\documentclass[preprint,journal]{vgtc}

\newcommand{\ie}{i.e.,~}
\newcommand{\etal}{\textit{et~al}. }
\newcommand{\etals}{\textit{et~al's}. }

\newcommand{\boldpara}[1]{{\textbf{#1.}}}

\long\def\symbolfootnote[#1]#2{\begingroup%
	\def\thefootnote{\fnsymbol{footnote}}\footnote[#1]{#2}\endgroup}

\usepackage{amsthm}
\newtheorem{theorem}{\sffamily Theorem}

\newtheorem{definition}[theorem]{\sffamily Definition}

\usepackage{amsmath}

\usepackage{amssymb}
\newcommand{\HG}{\mathsf{H}}
\newcommand{\V}{\mathsf{V}}
\newcommand{\E}{\mathsf{E}}
\newcommand{\C}{\mathcal{C}}

\usepackage[ruled, lined, noend, linesnumbered]{algorithm2e}

\SetCommentSty{mycommfont}

\usepackage{multirow}
\usepackage{colortbl}
\usepackage{threeparttable}
\usepackage{tabu}
\usepackage{booktabs}
\usepackage{mathptmx}

\preprinttext{To appear in IEEE Transactions on Visualization and Computer Graphics.}

\title{Structure-Aware Simplification for Hypergraph Visualization}

\author{
  \authororcid{Peter~Oliver}{0009-0002-5090-6057},
  \authororcid{Eugene~Zhang}{0000-0003-4752-3119}, \textit{Senior Member,~IEEE,}
  and \authororcid{Yue~Zhang}{0000-0002-8467-2781}, \textit{Member,~IEEE}
}

\authorfooter{
  \item
  Peter Oliver is with the School of Electrical Engineering and Computer Science, Oregon State University. E-mail: oliverpe@oregonstate.edu.
  \item
  Eugene Zhang is a Professor with the School of Electrical Engineering and Computer Science, Oregon State University. E-mail: zhange@eecs.oregonstate.edu.
  \item
  Yue Zhang is an Associate Professor with the School of Electrical Engineering and Computer Science, Oregon State University. E-mail: zhangyue@oregonstate.edu.

}

\abstract{
  Hypergraphs provide a natural way to represent polyadic relationships in network data. For large hypergraphs, it is often difficult to visually detect structures within the data. Recently, a scalable polygon-based visualization approach was developed allowing hypergraphs with thousands of hyperedges to be simplified and examined at different levels of detail. However, this approach is not guaranteed to eliminate all of the visual clutter caused by unavoidable overlaps. Furthermore, meaningful structures can be lost at simplified scales, making their interpretation unreliable. In this paper, we define hypergraph structures using the bipartite graph representation, allowing us to decompose the hypergraph into a union of structures including topological blocks, bridges, and branches, and to identify exactly where unavoidable overlaps must occur.
  We also introduce a set of topology preserving and topology altering atomic operations, enabling the preservation of important structures while reducing unavoidable overlaps to improve visual clarity and interpretability in simplified scales. We demonstrate our approach in several real-world applications.
}


\teaser{
  \centering
  \begin{subfigure}{0.32\textwidth}
    \centering
    \includegraphics[width=2in,angle=-90]{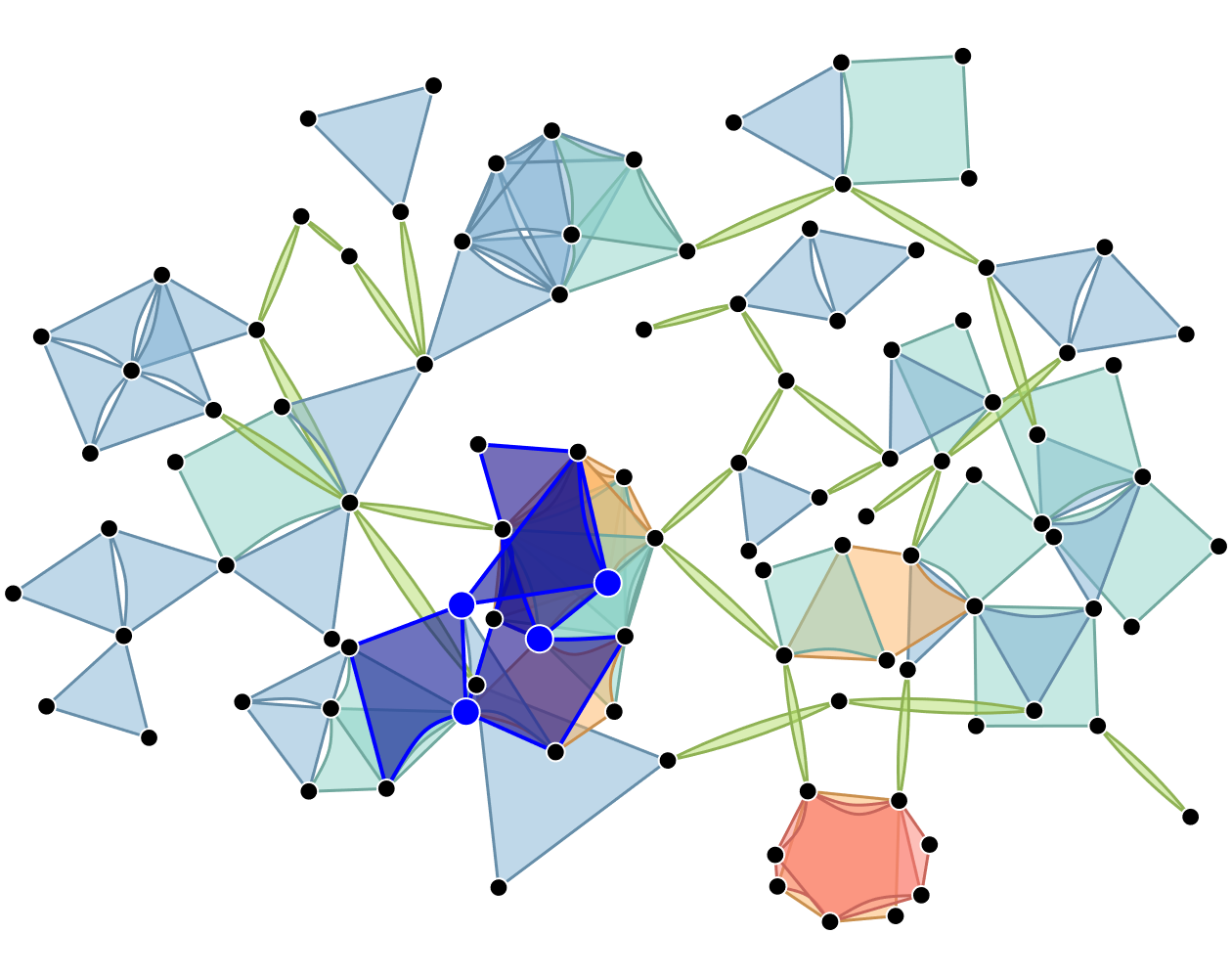}
    \caption{Input hypergraph}
  \end{subfigure}
  \begin{subfigure}{0.32\textwidth}
    \centering
    \includegraphics[width=2in,angle=-90]{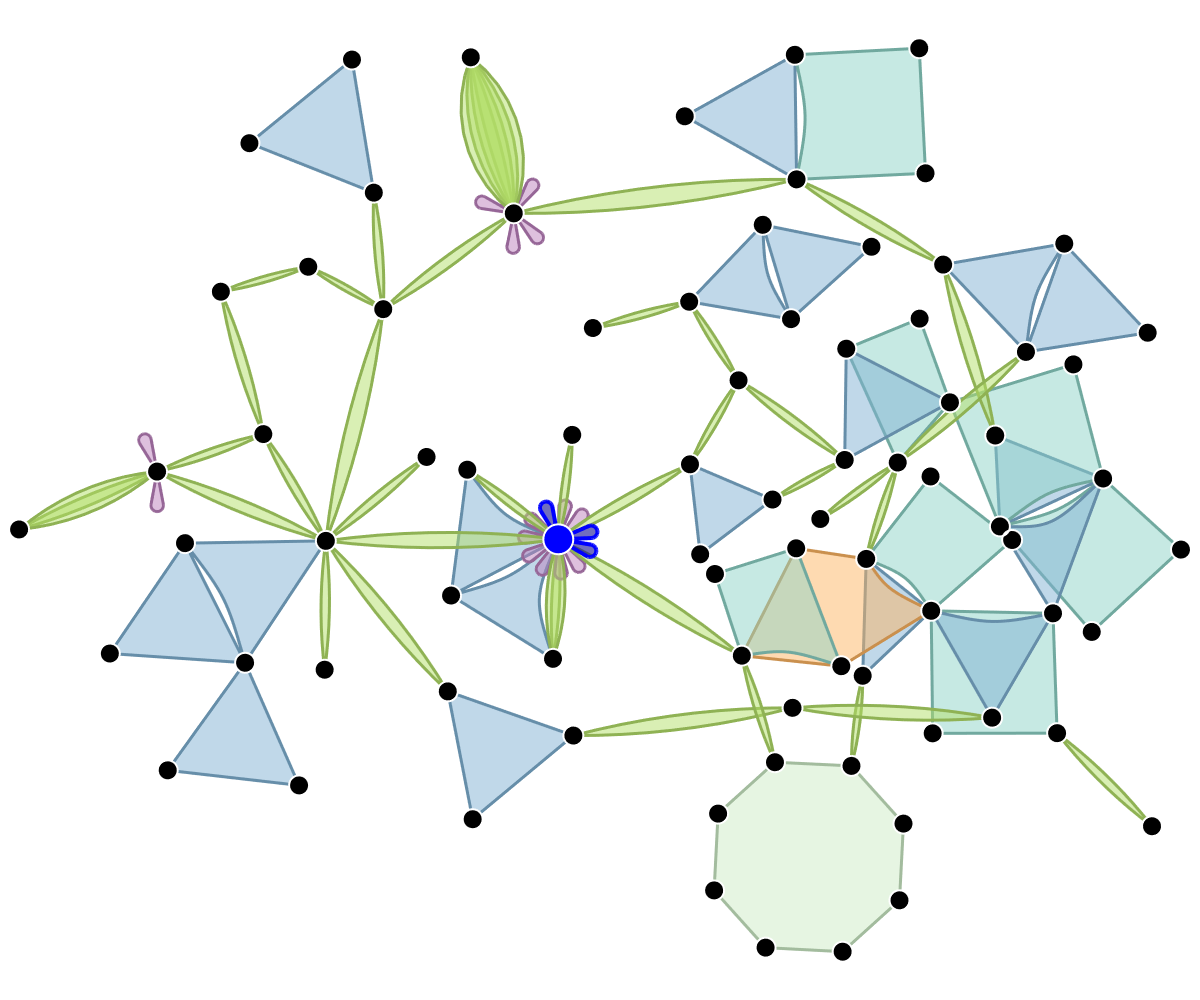}
	  \caption{Oliver \etal \cite{oliver2024scalable} simplified scale}
  \end{subfigure}
  \begin{subfigure}{0.32\textwidth}
    \centering
    \includegraphics[width=2in,angle=-90]{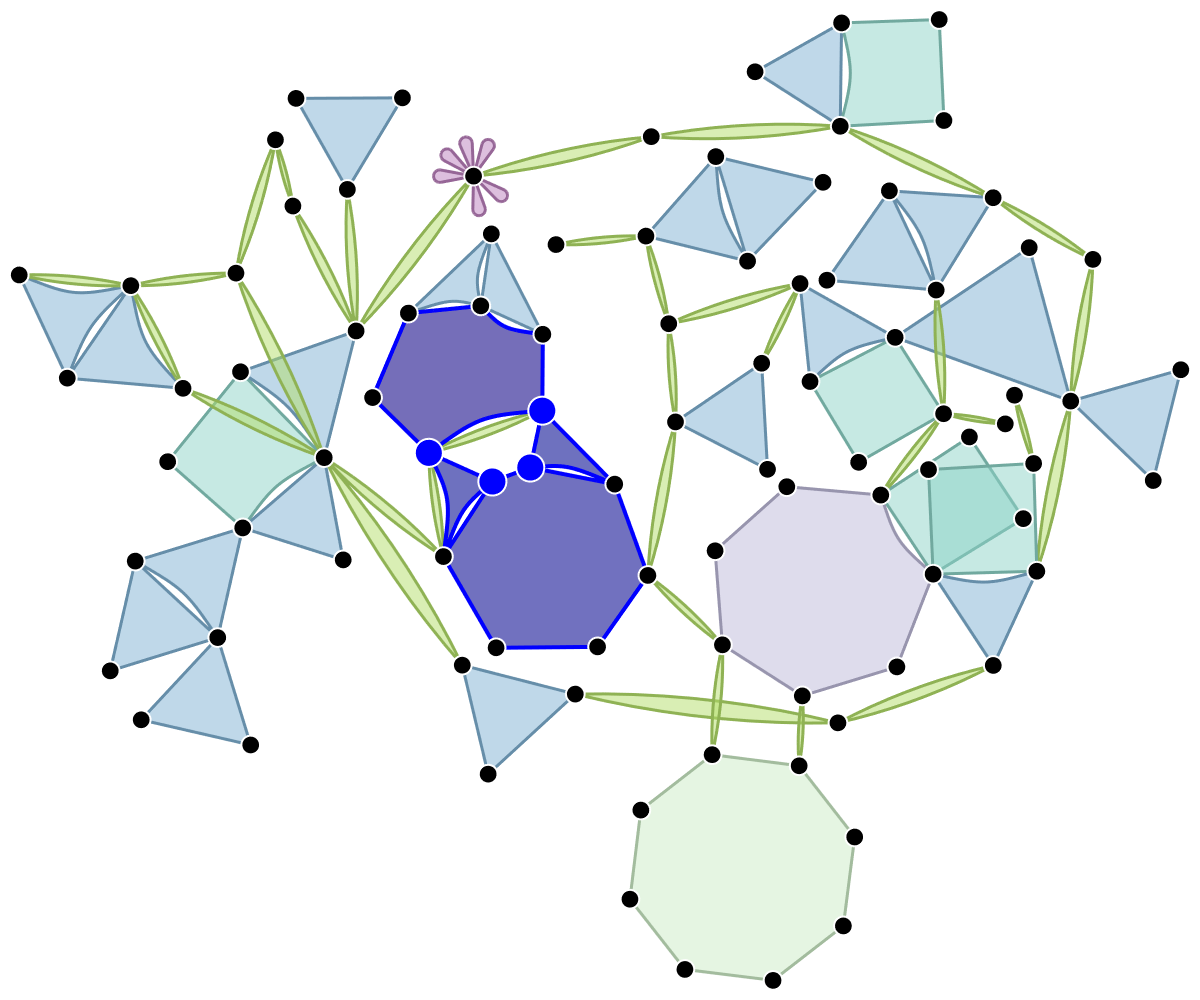}
    \caption{Our simplified scale}
  \end{subfigure}
  \caption{A hypergraph network representing friendships between 92 high school students in Marseilles~\cite{mastrandrea2015contact} is visualized using the polygon visualization metaphor. Visual clutter caused by overlapping polygons in (a) makes it difficult to identify cycle structures such as the one highlighted in blue. The simplification method of Oliver \etal~\cite{oliver2024scalable} shown in (b) reduces the polygon overlaps, but collapses the highlighted cycle into a single hypergraph vertex. This failure to preserve the cycle makes the interpretation of this intermediate scale less reliable. Our new structure-aware simplification method shown in (c) efficiently reduces polygon overlaps while preserving structures in the data such as the cycle. Our method enables multi-scale hypergraph visualizations where the simplified scales can be used to more easily identify meaningful structures in the data.}
  \label{fig:teaser}
}

\begin{document}

\maketitle

\section{Introduction}
\label{sec:introduction}

Polyadic relationships are omnipresent in network data with applications ranging from social networks, to paper authorship, and biology. Hypergraphs, as extensions to graphs, provide an ideal model for polyadic relationships where each relationship is represented as a {\em hyperedge}. The incident {\em vertices} of the hyperedge represent the entities in the underlying relationship. Any analysis of polyadic relationship data requires efficient visualization of the corresponding hypergraphs.

There have been recent advances in hypergraph visualization~\cite{Alsallakh:16}, with a focus on identifying an appropriate visual metaphor for hyperedges. Qu \etal~\cite{Qu:2017} propose the use of an $N$-sided polygon for each $N$-ary polyadic relationship in the data. In this \emph{polygon visualization metaphor}, the vertices of the polygons coincide with the entities in the $N$-ary relationship. In their follow-up research, Qu \etal~\cite{Qu:22} develop an optimization framework that generates high-quality polygon layouts for hypergraphs and their dual hypergraphs, where the roles of the vertices and hyperedges are switched. However, overlaps between polygons can make it challenging to differentiate individual polygons and correctly identify their shared vertices, especially for large datasets. \Cref{fig:teaser} (a) contains several examples. In addition, their optimization often becomes trapped at local minima in its objective function for datasets with more than a few hundred elements, leading to suboptimal layouts. To address this, Oliver \etal~\cite{oliver2024scalable} introduce a layout optimization framework in which the complexity of a hypergraph is iteratively reduced through a set of atomic simplification operations. Once the reduced hypergraph is sufficiently simple, an optimized layout is generated. From there, a sequence of inverse simplification operations is executed to gradually bring the complexity back to that of the original hypergraph. With each inverse operation, the layout is locally optimized and updated before continuing with the next inverse operation. This leads to a gain in layout quality and a reduction in the time required to produce such a layout. However, hypergraphs have structural features that can be lost at intermediate simplified scales, limiting their usefulness for multi-scale visualizations where significant structures should be recognizable in each scale. \Cref{fig:teaser} (b) shows an intermediate hypergraph scale created using Oliver et al.'s approach. Notice that the highlighted cycle structure in (a) is no longer present in (b). This cycle represents separation between friendship groups which is not easily visible in (a) due to polygon overlaps, and is completely removed in (b) where the friendship groups and the separation between them are collapsed into a single vertex. In addition, some visual clutter can persist in simplified scales, as shown in (b) by the remaining overlapping polygons, raising a fundamental question: is the visual clutter an artifact of the layout algorithm or is it unavoidable?

In this paper, we introduce a structure-aware approach to hypergraph simplification. We propose a novel decomposition of a hypergraph into an edge disjoint union of topological blocks, branches, and bridges. These structures can play an important role in interpreting hypergraph data. For example, the quantity and size of interweaving cycles in a social group hypergraph dataset can tell us how close-knit a particular group is. Alternatively, the presence of cycles in a contact tracing dataset indicates multiple possible transmission vectors for an infectious disease, namely, along either side of the cycle. Bridges and branches in a paper-author dataset can indicate links between research groups and help identify work that is on the outskirts of a research community. Our decomposition method also enables exact identification of areas in the hypergraph where unavoidable overlaps occur, which are a primary source of visual clutter in polygon visualizations. 

At the core of our decomposition and identification of unavoidable overlaps is the use of bipartite graphs as the foundation of hypergraph analysis. The bipartite graph representation replaces both the vertices and hyperedges of the hypergraph with graph nodes, using graph edges to indicate incidence relationships in the hypergraph. We define a \emph{topological block} as a maximal biconnected component in this bipartite graph. We present an efficient algorithm for computing a cycle basis for the bipartite graph, leading to a fast implementation of our decomposition. We also develop new theory showing how the bipartite cycles can be used to identify unavoidable overlaps in polygon visualizations. In particular, we define the \emph{entanglement index} as the ratio between the first Betti number (number of basis cycles) and the total number of elements in a topological block. We further show that unavoidable overlaps occur only in entangled topological blocks which we call \emph{forbidden clusters}. Powered by this analysis, we propose a new set of atomic simplification operations, inspired by the operations from \cite{oliver2024scalable}, using the nodes and edges of the bipartite graph representation as the fundamental units. We include both topology preserving and topology altering operations. A topology preserving operation does not affect the number of linearly independent bipartite cycles in the {\em cycle basis}, while a topology altering operation can decrease the number of independent cycles by one. Our hypergraph decomposition paired with the identified unavoidable overlaps allows us to determine exactly how and where the different types of operations should be applied to reduce visual clutter while preserving the most salient structures in the hypergraph. This also leads to a more efficient technique for producing planar simplified hypergraphs compared to \cite{oliver2024scalable}. We include three use cases with real-world datasets and provide our interpretations of the results.

\section{Previous Work}
\label{sec:prev}

Here we review recent work in network analysis and visualization that is most relevant to our paper.

\boldpara{Hypergraph Visualization}
Much of the recent work on hypergraph visualization has focused on identifying effective visual representations of hyperedges using matrices~\cite{Kim:2007,Sadana:2014,Lex:2014,valdivia2019analyzing}, regions~\cite{Rogers:08,simonetto2009fully,Stapleton:12,Micallef:14,Santamara:2010,Riche:2011,Alsallakh:2013,Arafat:17,simonetto2015simple}, metro lines~\cite{Wu:2020,Jacobsen:2021,Frank:2021}, and bipartite graphs~\cite{Stasko:07,Dork:12,Alsallakh:2013}. Matrix based methods enable spectral methods for analysis but do not as easily show structures in the data. Metro line visualizations can effectively display branching structures but are not ideal for representing topological structures like cycles. We use a bipartite graph representation for analysis, but for visualization, we adopt a region based approach of Qu \etal~\cite{Qu:22}. In their approach, each hyperedge is represented as a polygon drawn between the corresponding vertices embedded in the plane. By optimizing the positions of the vertices so that each polygon is as near to regular as possible, they generate high quality polygon layouts where the cardinality of a hyperedge can be identified through easily recognizable shapes.

In recent years, simplification has been used to reduce clutter and enhance readability in hypergraph visualizations. Oliver \etal~\cite{oliver2024scalable} build on the layout optimization framework of~\cite{Qu:22}, developing a multi-scale optimization method that is more effective in reducing polygon overlaps in the layout. Recognizing the potential of using simplified scales to study hypergraph structures, we build on the simplification scheme of~\cite{oliver2024scalable}, augmenting their statistic based approach with a new structure-aware implementation that can precisely identify sources of unavoidable overlap. Zhou \etal~\cite{zhou2023simplification} apply persistent homology techniques to compute a \emph{barcode} for a chosen graph representation of the input hypergraph, which is used to guide merging operations. Their aim is to reduce the size of the data to create more compact and readable visualizations. In contrast, Oliver \etal~\cite{oliver2024scalable} apply simplifications directly to the hypergraph vertices and hyperedges, guided by a customizable priority measure. They aim to provide scalable visualization techniques for very large datasets. Neither approach directly addresses visual clutter caused by unavoidable overlaps. To achieve simultaneous vertex and hyperedge simplification, we apply simplifications to the bipartite graph representation.

\boldpara{Hypergraph Structures} Recent work from Fan \etal~\cite{Fan2021cycles} focuses on cycles in graph data with the aim to improve quantification on important nodes. However, such analysis is not focused on bipartite graphs which correspond to hypergraphs. Considerable effort has been made to describe connectedness in various mathematical representations. In addition to prominent work by Erd{\"o}s~\cite{erdos1959circuits} and Berge~\cite{berge1973graphs} on hypergraph combinatorics, Aksoy \etal~\cite{aksoy2020hypernetwork} introduce concepts of high-order $s$-walks and $s$-paths which have concepts of both length and width. Such structures are powerful tools to analyze the connectivity of a hypergraph. To our knowledge, we are the first to explore the decomposition of a hypergraph into its constituent structures to guide structure-aware simplification and identify unavoidable visual clutter.

\section{Background}
\label{sec:definitions}

\setlength{\abovedisplayskip}{4pt}
\setlength{\belowdisplayskip}{0pt}

\begin{figure}[bpt]
  \centering
  \subfloat[][Primal hypergraph $\HG$]{\includegraphics[height=1.25in,]{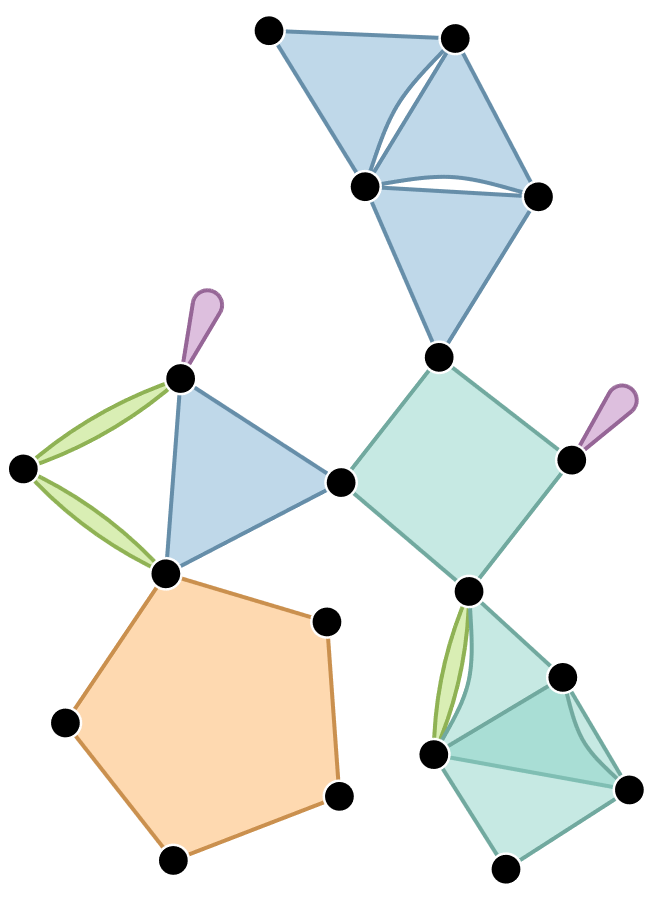}} \hspace{0.25in}
  \subfloat[][Bipartite representation]{\includegraphics[height=1.25in,]{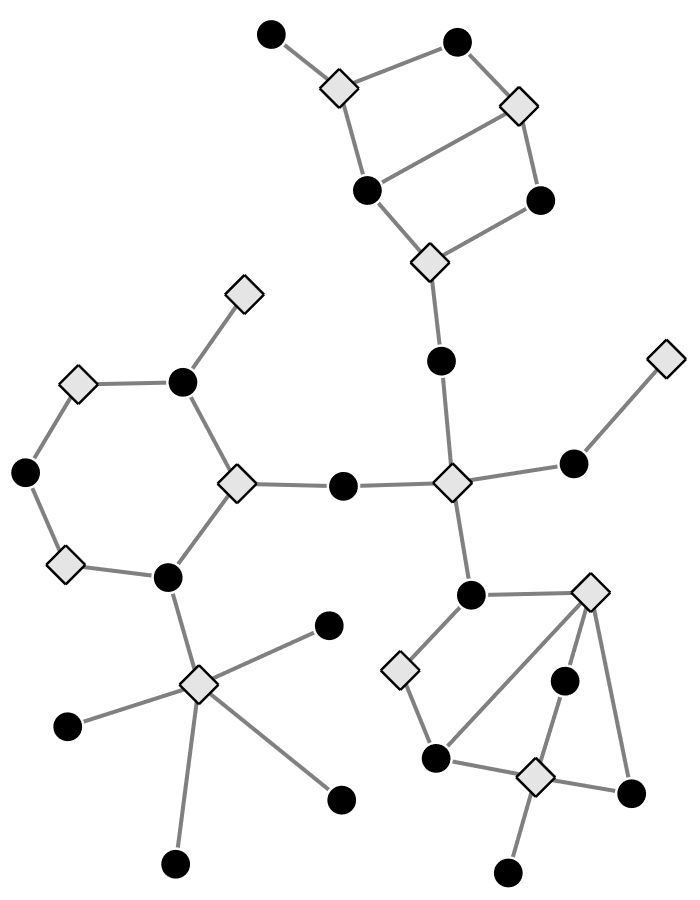}} \hspace{0.25in}
  \subfloat[][Dual Hypergraph $\HG^*$]{\includegraphics[height=1.25in,]{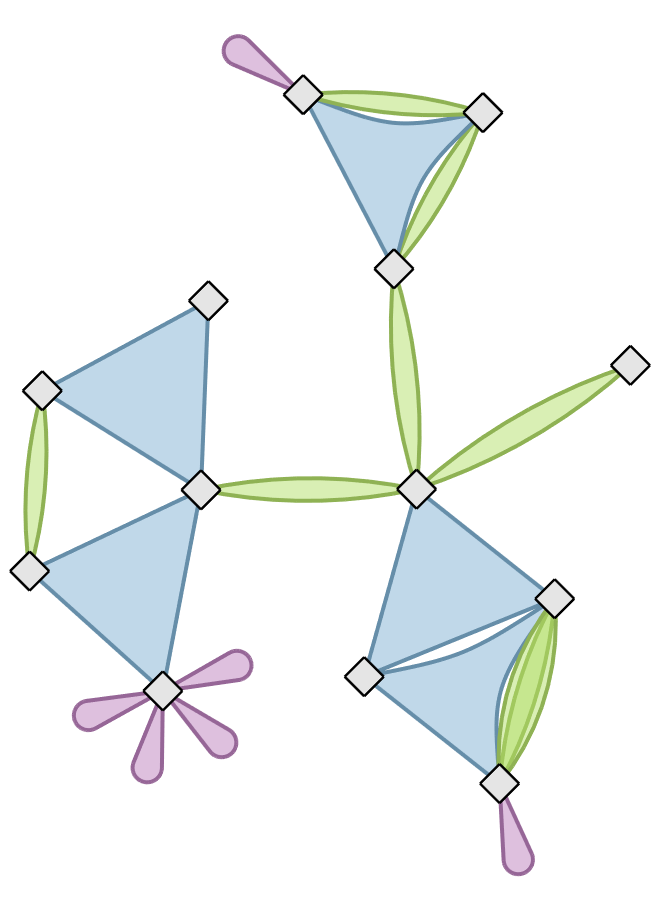}}
  \caption{The bipartite representations (b) of the primal hypergraph (a) and dual hypergraph (b) are identical. We use a black dot to represent a node in the primal hypergraph and a gray diamond to represent a node in the dual hypergraph.}
  \label{fig:bipartite}
\end{figure}

A hypergraph $\HG=\langle \V,\E \rangle$ on a finite set of vertices $\V$ is defined by a family of hyperedges $\E$. A hyperedge $e \in \E$ contains a non-empty subset of vertices $\V_e \subseteq \V$ which are \emph{incident} to $e$ and \emph{adjacent} to each other. Similarly, a vertex $v \in  \V$ is contained by a subset of hyperedges $\E_v \subseteq \E$ which are \emph{incident} to $v$ and \emph{adjacent} to each other. Following Oliver \etal~\cite{oliver2024scalable}, let $\E_e$ denote the set of hyperedges adjacent to $e$ and $\V_v$ the set of vertices adjacent to $v$. Consistent with \cite{Qu:22}, we define the \emph{degree} of a vertex $v$ as $\text{deg}(v) = |\E_v|$ and the \emph{cardinality} of a hyperedge $e$ as $\text{card}(e) = |\V_e|$.

The \emph{dual hypergraph} $\HG^*=\langle \V^*,\E^*\rangle$ of $\HG$ is obtained by switching the roles of vertices and hyperedges in $\HG$ (\Cref{fig:bipartite} (c)). We refer to the original hypergraph $\HG$ as the \emph{primal hypergraph}. More precisely, each element $v \in \V$ corresponds to a unique element $v^* \in \E^*$ and each element $e \in \E$ corresponds to a unique element $e^* \in \V^*$. The incidence and adjacency relationships of corresponding elements in the primal and dual hypergraphs are identical. This means that the primal and dual hypergraphs share combinatorial features and properties such as linearity and planarity~\cite{berge1973graphs,oliver2024scalable}.

For $\HG=\langle \V,\E \rangle$, the bipartite representation (also called the K{\"o}nig representation) ${G(\HG)=(X \cup Y,D)}$ is a bipartite graph with nodes $x \in X$ for every vertex in $\V$ and nodes $y \in Y$ for every hyperedge in $\E$ (\Cref{fig:bipartite} (b)). There is a one-to-one correspondence between $\V$ and $X$, so without ambiguity, we write $X=\V$. Similarly, there is a one-to-one correspondence between $Y$ and $\E$, i.e. every node in $Y$ corresponds precisely to one hyperedge in $\HG$. Again, we write $Y=\E$. Vertices $x \in X$ and $y \in Y$ form an edge $(x,y) \in D$ if and only if $x$ is part of the hyperedge associated with $y$. We can also define the bipartite representation in terms of the dual hypergraph $\HG^*$, resulting in an equivalent graph $G^*$. Thus, we identify $G$ and $G^*$ as one graph. A node in $G$ that is a vertex in $\HG$ we refer to as a {\em primal node} and is drawn as a black dot in our visualizations (\Cref{fig:bipartite}). A node in $G$ that corresponds to a hyperedge in $\HG$ (or equivalently a vertex in the dual hypergraph) is referred to as {\em dual node} and is drawn as a grey diamond. Notice that an edge in $D$ must link a primal node to a dual node, thus $G$ is bipartite.

\begin{figure}
  \centering
  \includegraphics[height=0.7in]{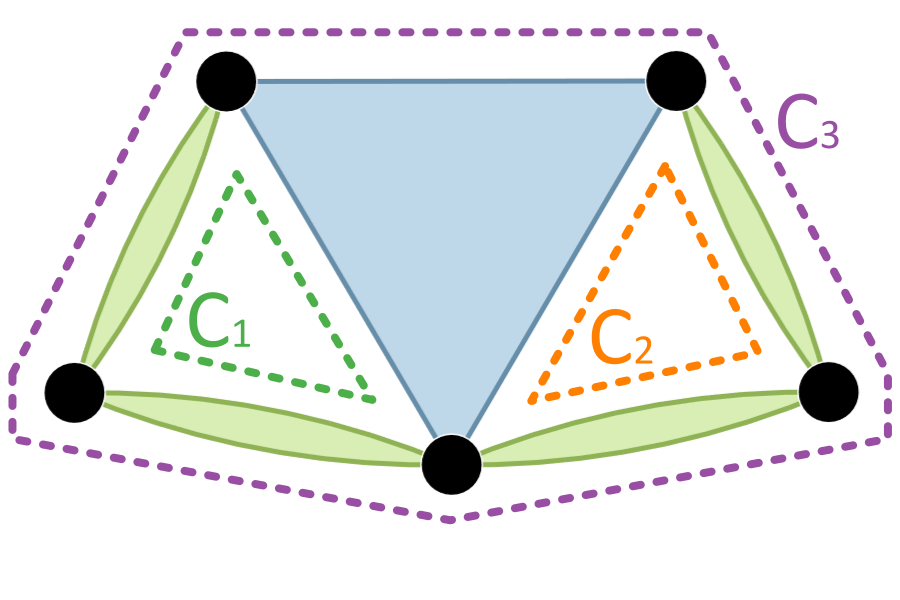} \hspace{0.1in}
  \includegraphics[height=0.7in]{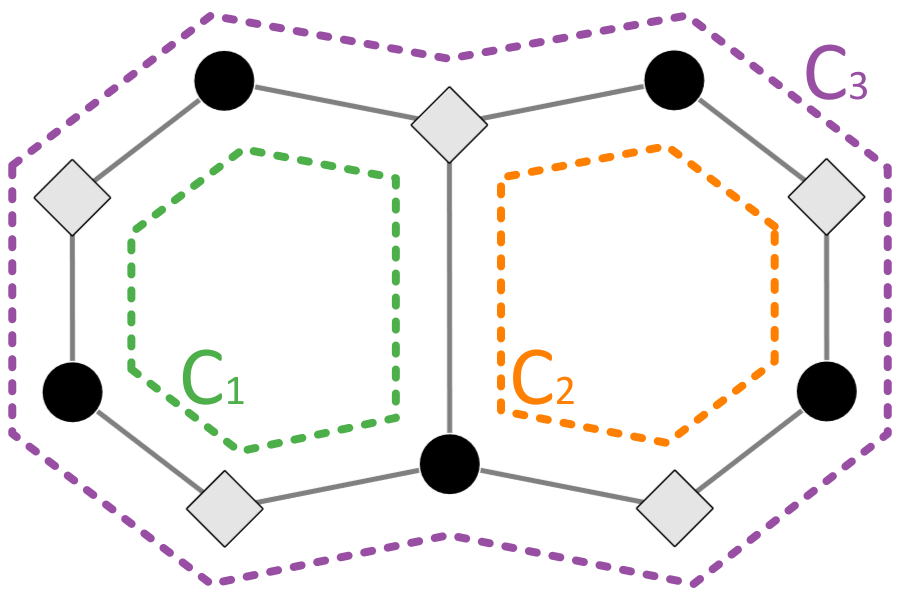} \hspace{0.1in}
  \includegraphics[height=0.7in]{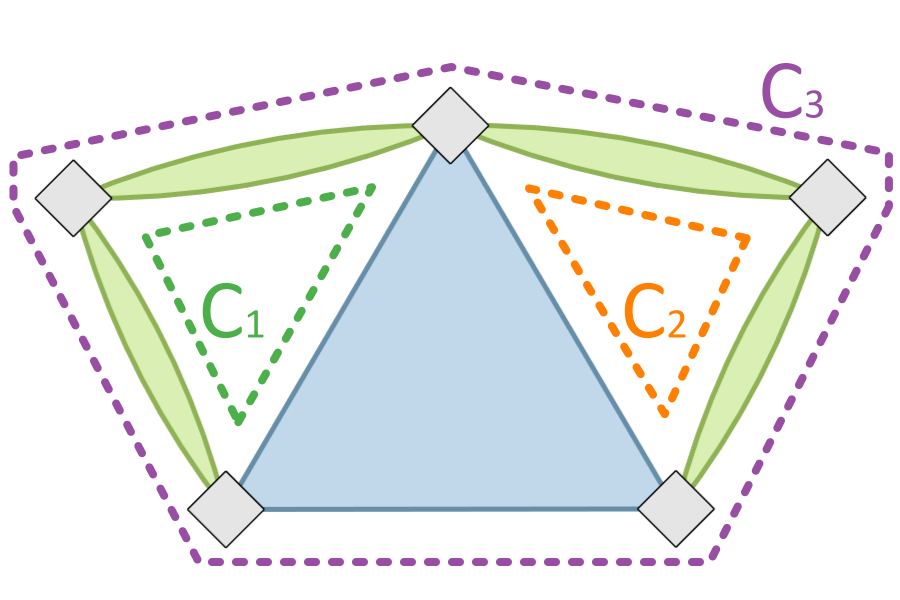}
  \caption{The three simple cycles $C_1,C_2,C_3$ of a bipartite graph (center) are highlighted by the green, orange, and purple dotted lines. The same cycles are highlighted in the matching primal hypergraph (left) and dual hypergraph (right). Cycles $C_1$ and $C_2$ can be combined to form $C_3$.}
  \label{fig:loops}
\end{figure}

Bretto ~\cite{bretto2013hypergraph} defines a \textit{path} $P$ in $\HG$ from $u$ to $v$, $u,v \in \V$, as an alternating sequence \mbox{$P = \langle u=x_1,e_1,x_2,e_2,\dots,x_n,e_n,x_{n+1} = v \rangle$} where
\begin{itemize}[noitemsep]
  \item $x_1,x_2,\dots,x_n \in V$ are distinct vertices,
  \item $e_1,e_2,\dots,e_n \in E$ are distinct hyperedges,
  \item $x_i,x_{i+1} \in e_i$ for $1 \leq i \leq n$.
\end{itemize}
$\HG$ is \textit{connected} if every distinct pair of vertices is linked by some path $P$. If $u = x_1 = x_{n+1} = v$, then $P$ is a \textit{cycle}. \Cref{fig:loops} outlines three such cycles in the primal, bipartite, and dual visualizations. This definition of a hypergraph cycle is consistent with Berge ~\cite{berge1973graphs}, and is analogous to a simple cycle in a graph where none of the vertices is repeated. We assume that all graph and hypergraph cycles discussed in the remainder of the paper are simple.

The simple cycles of the bipartite representation $G$ live within \emph{2-connected components}. Any vertex in $G$ whose removal makes $G$ disconnected, or increases the number of connected components, is called an \emph{articulation vertex}. A connected graph that has no articulation vertices is called 2-connected. A graph can contain multiple 2-connected subgraphs. A maximal connected subgraph in $G$ containing no articulation vertices is also called a \emph{block}. A block on more than two vertices is called a 2-connected (or biconnected) component. Blocks can consist of an isolated node or a single edge, while the smallest 2-connected component is a triangle. Cycles cannot exist outside of a 2-connected component without containing duplicate nodes or edges.

\section{Topology Guided Decomposition}
\label{sec:decomposition}

While past analysis of hypergraphs has focused on local behaviors, such as the cardinality of a hyperedge or the intersections between hyperedges, hypergraphs have intricate structures that are global in nature. For example, cycles involving multiple vertices and hyperedges can exist, and a pair of cycles can be connected by a single path without which they would be disconnected. In this section, we define a number of hypergraph structures and develop a decomposition for hypergraphs into a union of these features. We also provide an efficient algorithm to compute the decomposition. The structures lead to a number of atomic simplification operations, enabling multi-scale representations of hypergraphs in which important features are preserved at simplified scales, providing meaningful visual interpretations.

\subsection{Topological Blocks, Bridges, and Branches}

Notice that a path $P$ in $\HG$ induces a subgraph in the bipartite representation which is a graph path between alternating primal and dual nodes in $G$. If $P$ is a cycle, the subgraph in $G$ induced by $P$ is also a cycle. The reverse is also true if we relax the definition of a hypergraph cycle to allow the starting and ending points to be hyperedges, \ie the vertices of a path in $G$ induce hypergraph paths in $\HG$ and $\HG^*$. Thus, we can identify cycles in $\HG$ and $\HG^*$ using the cycles of $G$ and vice versa (\Cref{fig:loops}). In this paper, we treat all hypergraphs, and by extension their bipartite representations, as unweighted. Thus, the weight of any path in $G$ is simply the length of the path. We use the bipartite path length to describe the lengths of cycles in the primal and dual hypergraphs as well, \ie a hypergraph cycle containing $n$ vertices has an overall length of $2n$ since it must also contain $n$ hyperedges. The smallest possible cycle contains two vertices and two hyperedges, has a length of 4, and is represented in $G$ as the complete bipartite graph $K_{2,2}$. We refer to these as \emph{minimal cycles}.

We also consider the topology of $\HG$ in terms of the topology of the bipartite graph $G$. Using the language of homology~\cite{Kaczynski:2004}, the topology of $G(\HG)$ can be measured in terms of its {\em Betti numbers}. The zeroth Betti number, $B_0$, is the number of connected components in $G$. In our datasets, all of the hypergraphs are connected, so $B_0=1$. On the other hand, the first Betti number, $B_1$, is the number of independent cycles in the graph. A cycle can be considered as the combination of two or more other cycles as shown in \Cref{fig:loops}. In this case, only two of the three cycles are mutually independent, \ie $B_1=2$.

In fact, there exists a {\em cycle basis} $\C=\{C_1, \dots, C_p\}$ where each $C_i$ ($1\le i\le p$) is an independent cycle in $G$, and any cycle of $G$ not in $\C$ can be written as a combination of two or more cycles in $\C$. Here, cycles are combined using the symmetric difference operator. Such a basis $\C$ spans the \emph{cycle space} of $G$. Furthermore, $\C-\{C_i\}$ for any index $i$ no longer spans the cycle space so it is not a cycle basis. There are a number of choices for a cycle basis of $G$, however, the number of cycles $p$ in any basis is given by the first Betti number $B_1$. A common problem in graph theory is to find a cycle basis such that the sum of all basis cycle weights is minimum. Such a basis is called a \emph{minimal cycle basis}. Horton~\cite{horton1987basis} prove that a minimum cycle basis consists only of \emph{tight} cycles. A cycle $C$ is \emph{tight} if the shortest path between every pair of nodes in $C$ is a sub-sequence of $C$. In \cref{fig:loops}, $C_1$ and $C_2$ are tight, but $C_3$ is not tight because it does not contain the shortest path between the middle primal and dual nodes. Furthermore, each of the pairings $\{C_1,C_2\}$, $\{C_1,C_3\}$, and $\{C_2,C_3\}$ define a cycle basis, but $\{C_1,C_2\}$ is the only minimum cycle basis and it contains tight cycles.

\begin{figure}[tbp]
  \centering
  \subfloat[][Block decomposition of $G$]{\includegraphics[height=1.4in]{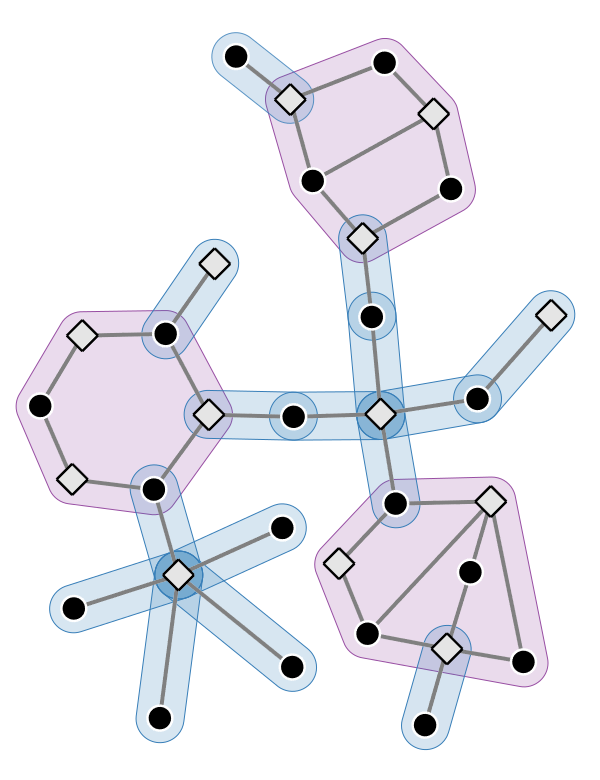}}
  \subfloat[][Topological decomposition $G$]{\includegraphics[height=1.4in]{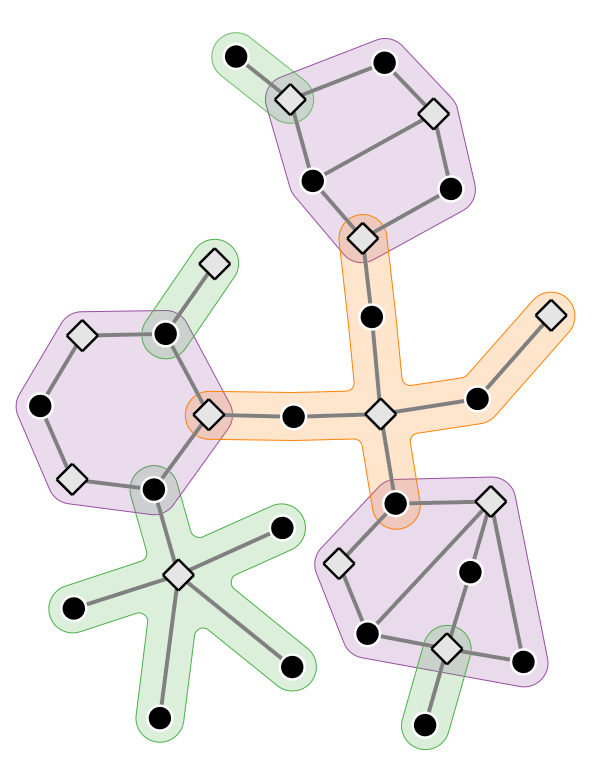}}
  \subfloat[][Hypergraph structures]{\includegraphics[height=1.4in]{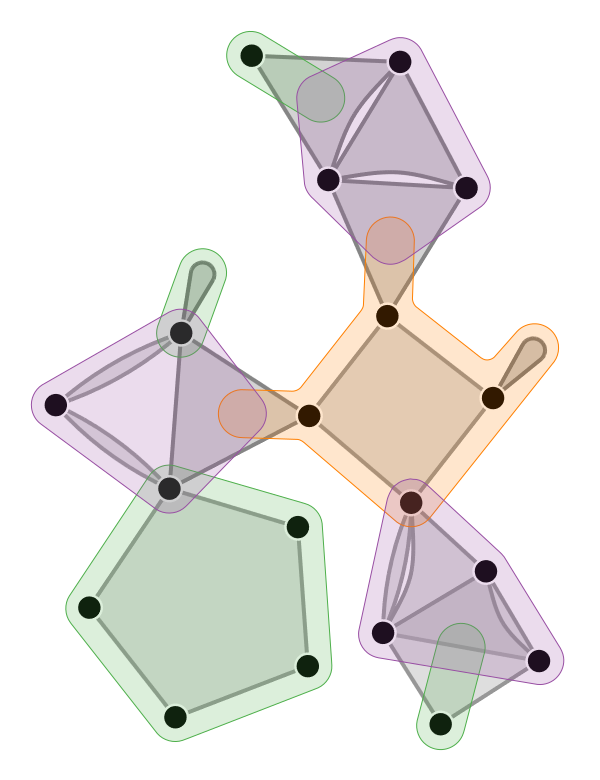}}
  \caption{We use our block decomposition (a) to generate a topological decomposition (b) of the bipartite graph representation for the hypergraph in \Cref{fig:bipartite}. In (a), the blue bubbles indicate single edge blocks and the purple bubbles multi-edge blocks. This leads to a number of extracted structures in (c) including topological blocks (purple bubbles), bridges (orange bubbles), and branches (green bubbles).}
  \label{fig:topo_decomp}
\end{figure}

\begin{figure*}[tbp]
  \centering
  \captionsetup[subfigure]{labelformat=empty}
  \subfloat[][(a1)]{\includegraphics[height=0.65in]{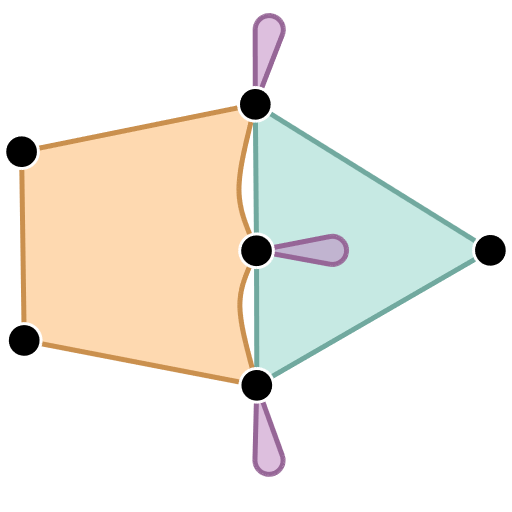}} \hspace{0.05in}
  \subfloat[][(a2)]{\includegraphics[height=0.65in]{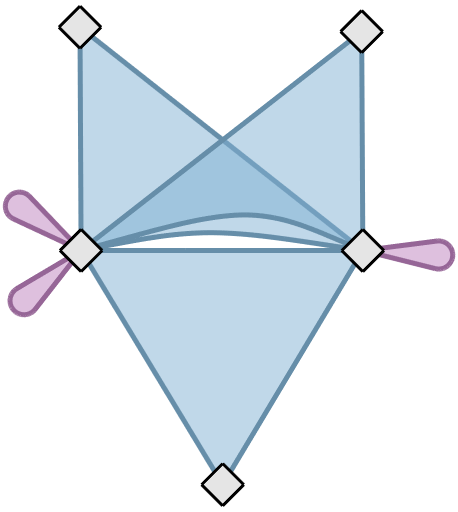}} \hspace{0.05in}
  \subfloat[][(a3)]{\includegraphics[height=0.65in]{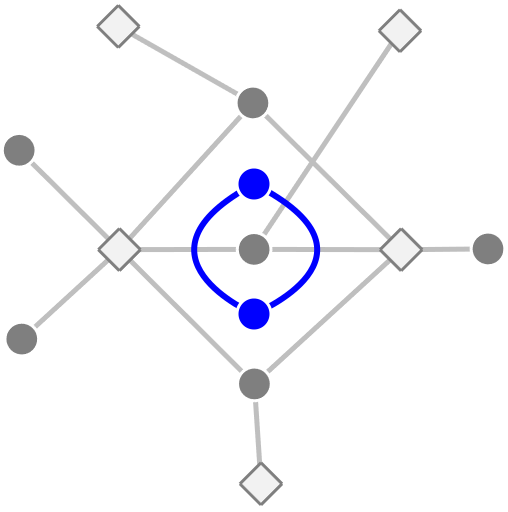}} \hspace{0.25in}
  \subfloat[][(b1)]{\includegraphics[height=0.65in]{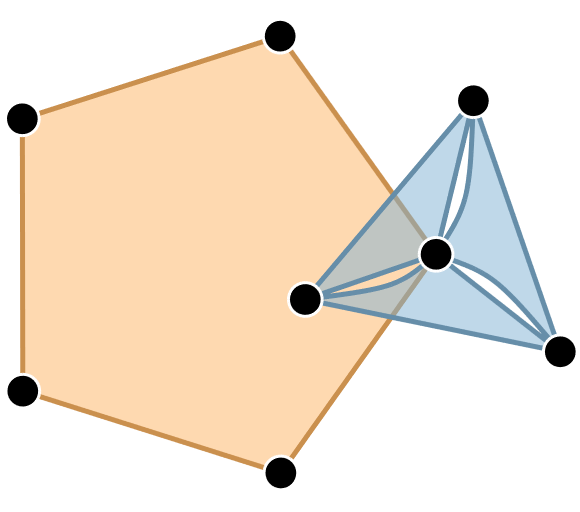}} \hspace{0.1in}
  \subfloat[][(b2)]{\includegraphics[height=0.65in]{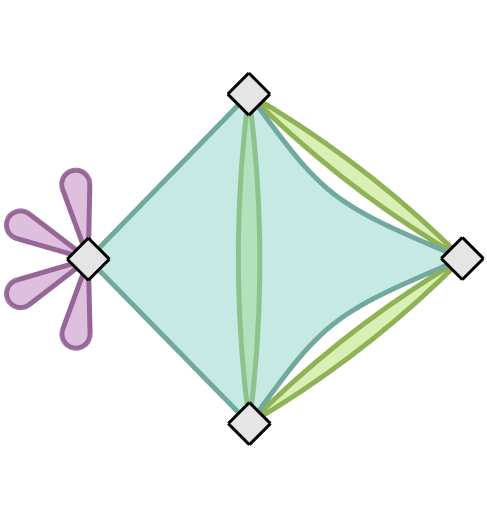}} \hspace{0.05in}
  \subfloat[][(b3)]{\includegraphics[height=0.65in]{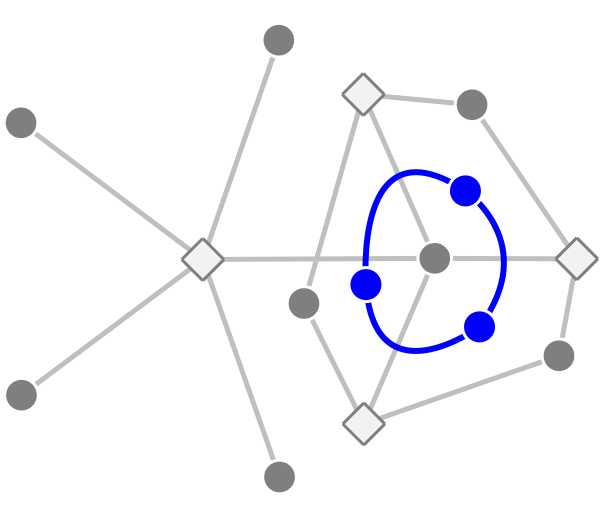}} \hspace{0.25in}
  \subfloat[][(c1)]{\includegraphics[height=0.55in]{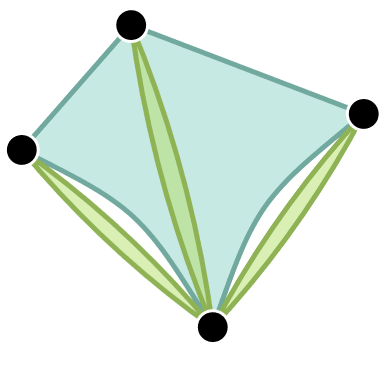}} \hspace{0.05in}
  \subfloat[][(c2)]{\includegraphics[height=0.55in]{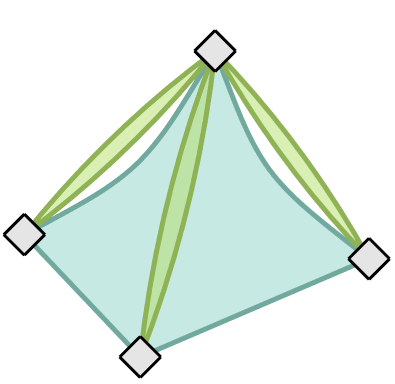}} \hspace{0.1in}
  \subfloat[][(c3)]{\includegraphics[height=0.55in]{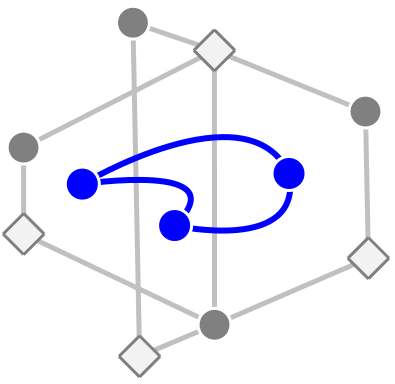}}
  \caption{The forbidden sub-hypergraphs of polygon hypergraph drawings: (a1) 3-adjacent hyperedge bundle of 2 hyperedges, (a2) 2-adjacent hyperedge bundle of 3 hyperedges, (b1) strangled vertex cycle variant, (b2) strangled hyperedge cycle variant, and (c1, c2) strangled vertex and hyperedge star variant. Notice that (a2) is the dual of (a1), (b2) is the dual of (b1), and (c2) is the dual of (c1). The cycle adjacency graph for each primal-dual pair is drawn in blue over the corresponding bipartite graph representation in (a3), (b3), and (c3).}
  \label{fig:forbidden}
\end{figure*}

The topological structure of $\HG$ naturally leads to a decomposition of both the hypergraph and the bipartite representation $G$. Let $\C$ be a cycle basis for $\HG$. Within $\C$, we consider two basis cycles $C_1$ and $C_2$ to be connected if they share a common edge in $G$, \ie they share at least one primal node and one dual node. Suppose that $\C$ has $q$ connected components $\{T_1,\dots,T_q\}$. We refer to each connected component $T_i$ as a {\em topological block}. Let $K = G-\C$. Then $K=\bigcup_{i=1}^r K_i$ is the disjoint union of $K_i$'s where each subgraph $K_i$ is a tree. This is illustrated in \Cref{fig:topo_decomp} (a) where the purple bubbles indicate extracted topological blocks and the blue bubbles indicate the remaining tree subgraphs. Each $K_i$ can be connected to one or more topological blocks. For each block $T_j$ connected to $K_i$, we call the node $x_j \in T_j$ a \emph{root} of $K_i$. Note that $K_i$ cannot have more than one root in a single topological block, otherwise, it would define a new cycle. Similarly, a pair of topological blocks cannot be connected by more than one $K_i$. If a tree $K_i$ is connected to only one topological block, having only one root node, we refer to it as a {\em branch}. \Cref{fig:topo_decomp} (b,c) show that tree structures can be rooted at either primal or dual bipartite nodes. On the other hand, if $K_i$ is connected to two or more topological blocks, thus having two or more roots, we refer to it as a {\em bridge}. \Cref{fig:topo_decomp} (b,c) shows a bridge highlighted with an orange bubble that is connected to three topological blocks, having one primal node root and two dual node roots. This example also shows that a bridge, like a branch, can contain leaf nodes in $G$. We use the roots of a tree $K_i$ to distinguish between bridges and branches instead of using leaf nodes, although we note that a branch necessarily contains one or more leaf nodes while a bridge can have no leaf nodes. Thus, our bipartite graph is an edge-disjoint union of basis cycles in $\C$ which form the topological blocks, the set of all branches, and the set of all bridges (\Cref{fig:topo_decomp} (b,c)). We refer to this as the \emph{topological decomposition} $D$ of the bipartite graph $G$ and corresponding hypergraph $\HG$.

\subsection{Decomposition Algorithm}
\label{sec:algorithm}

To find the topological decomposition $D$ of the bipartite graph $G(\HG)$, we first compute a more granular decomposition $D'$ based on the blocks of $G$. Recall that a \emph{block} in $G$ is a maximal connected subgraph that has no articulation nodes. Following the classic algorithm of Hopcroft and Tarjan~\cite{hopcroft1973Algorithm}, we use a depth first search to extract all the blocks of $G$. The result is an edge-disjoint decomposition $D'$ of $G$ into a set of blocks. Within $D'$, we have two types of blocks: those consisting of a single edge in $G$, and those consisting of multiple edges and nodes in $G$. These are shown using blue and purple bubbles respectively in \Cref{fig:topo_decomp} (a). Notice that the multi-edge blocks are exactly the topological blocks from our cycle-based decomposition $D$. That is, each multi-edge block consists of a subset of connected cycles in a cycle basis $\C$ of $G$. Furthermore, a single-edge block cannot belong to any cycle in $G$ since both of its endpoints are articulation nodes in $G$. This means that every single-edge block in $D'$ is contained within some bridge or branch structure in $D$. We construct the bridges and branches of $D$ by finding connected components of the single-edge blocks in $D'$.

In practice, computing the decomposition $D'$ and using it to extract the topological blocks, bridges, and branches of $D$ can be performed within a single depth-first search. We augment the algorithm of Hopcroft and Tarjan~\cite{hopcroft1973Algorithm} to keep track of the multi-edge blocks in $D'$ as well as the connected components of single-edge blocks in $D'$. For an arbitrary hypergraph $\HG$ and its bipartite representation $G$, our algorithm produces a topological decomposition in $O(|V(G)|+|E(G)|)$ time.

As an additional step, we also extract a cycle basis $\C$ associated with the topological decomposition $D$. Instead of searching all of $G$ for cycles, we compute a cycle basis for each topological block $T \in D$ individually. Since $G$ is unweighted, we use a pair of nested breadth-first searches to find a basis of linearly independent tight cycles in $T$. Our algorithm is inspired by Gashler and Martinez~\cite{gashler2012cyclecut} who use a similar algorithm to find topological holes in manifold learning datasets. The first breadth-first search builds up a subgraph $S$ of traversed edges in $T$. When the first search discovers a back edge $(x,y)\in E(T)$ connecting to a previously visited node $y$, a new breadth-first search on $S$ is started at $x$. Once this second breadth-first search reaches $y$, the search path from $x$ to $y$ along with the edge $(x,y)$ is saved as a new basis cycle. Implementing the second search as a breadth-first search ensures that we find the shortest path in $T$ from $x$ to $y$ apart from the edge $(x,y)$. This means that every extracted cycle is a tight cycle. We provide pseudocode for this algorithm in the supplementary material \Cref{apx:algorithm}.

\section{Planarity}
\label{sec:planarity}

In a topological decomposition of hypergraph $\HG$, the set of bridges and branches are planar both in the bipartite representation $G$ and in the polygon representation of $\HG$. This is because the bridges and branches all have a tree structure. Consequently, any non-planarity in $G$ and any unavoidable overlaps in $\HG$ must occur within the topological blocks. Note that unavoidable overlaps are a major source of visual clutter in hypergraph visualizations. However, a topological block $T$ is not guaranteed to contain unavoidable overlaps. A set of connected cycles can have a planar representation if they are not entangled. 

Consider one such topological block $T$, which is connected by definition and contains a subset of the basis cycles in $\C$. Thus, the Betti numbers $B_0(T)=1$ and $B_1(T)>0$. Let $\chi(T)=|V(T)|-|E(T)|$ denote the Euler characteristic of $T$. We also know that $\chi(T)=B_0(T)-B_1(T)$. Since $B_0(T)=1$, we have $B_1(T)=1+|E(T)|-|V(T)|$. That is, the number of independent cycles in a topological block is directly related to the difference in the number of edges and the number of vertices in the block. The more cycles in the block, the more likely that $T$ is entangled. We thus define the {\em entanglement index} of $T$ as $\eta(T)=\frac{B_1(T)}{|V(T)|}$. Subgraphs that are trees, such as our bridge and branch structures, do not contain any cycles and have an entanglement index of zero.

The entanglement index can be used to determine which topological blocks likely contain unavoidable overlaps. However, within an entangled topological block, the unavoidable overlaps may only occur among a small subset of the cycles in the block. In this section, we develop theory on the configurations that cause unavoidable overlaps using the language of our topological decomposition.

\begin{figure*}[btp]
  \centering
  \subfloat[][Primal hypergraph]{\includegraphics[height=1.5in]{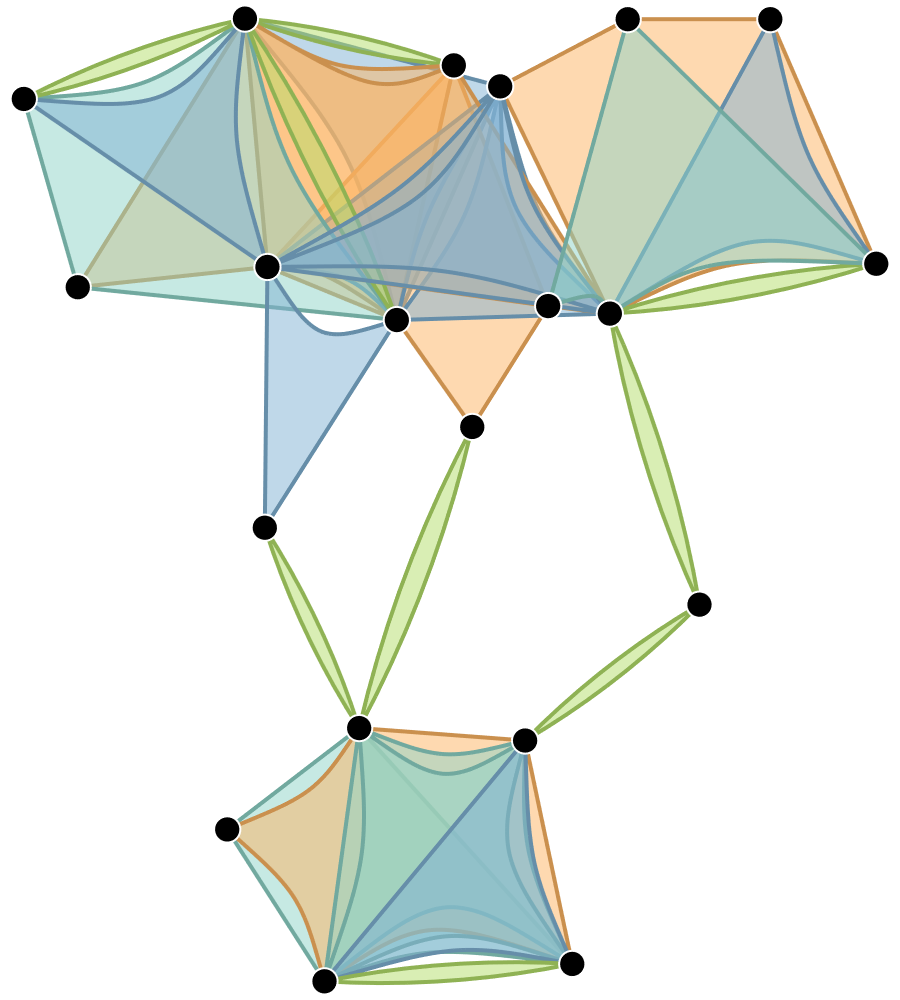}} \hspace{0.25in}
  \subfloat[][Bipartite representation]{\includegraphics[height=1.5in]{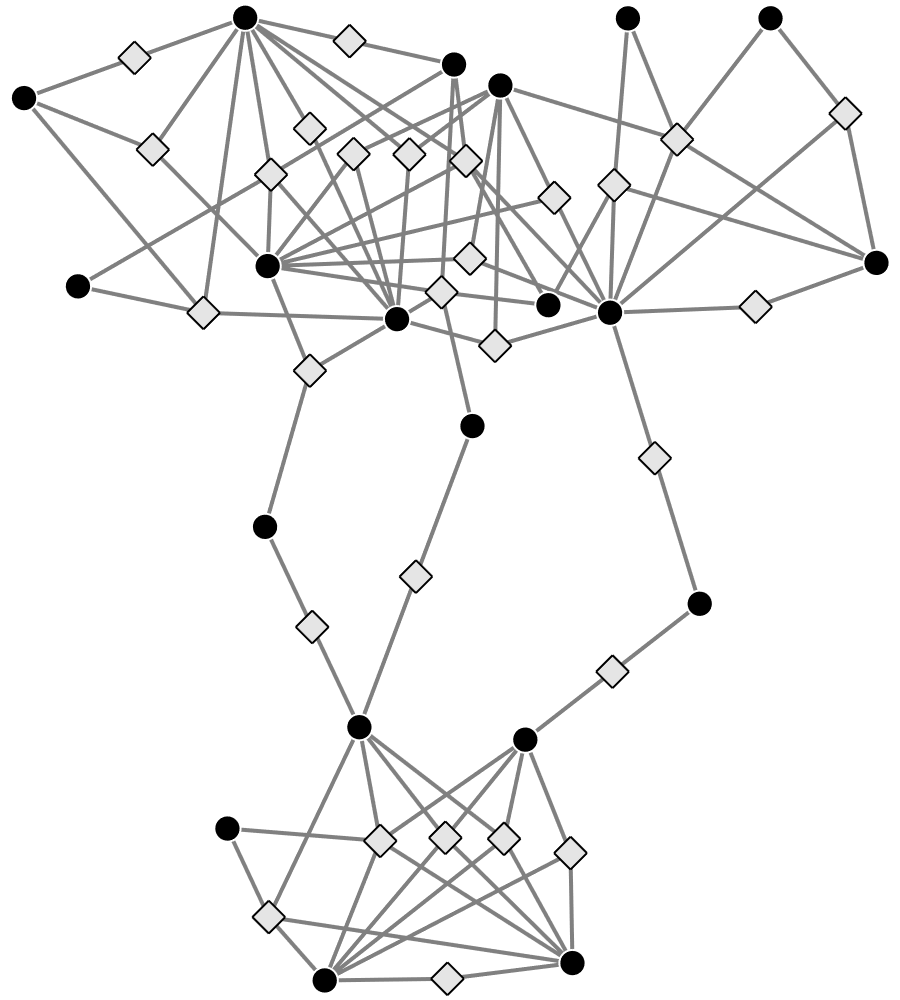}} \hspace{0.25in}
  \subfloat[][Cycle adjacency graph]{\includegraphics[height=1.5in]{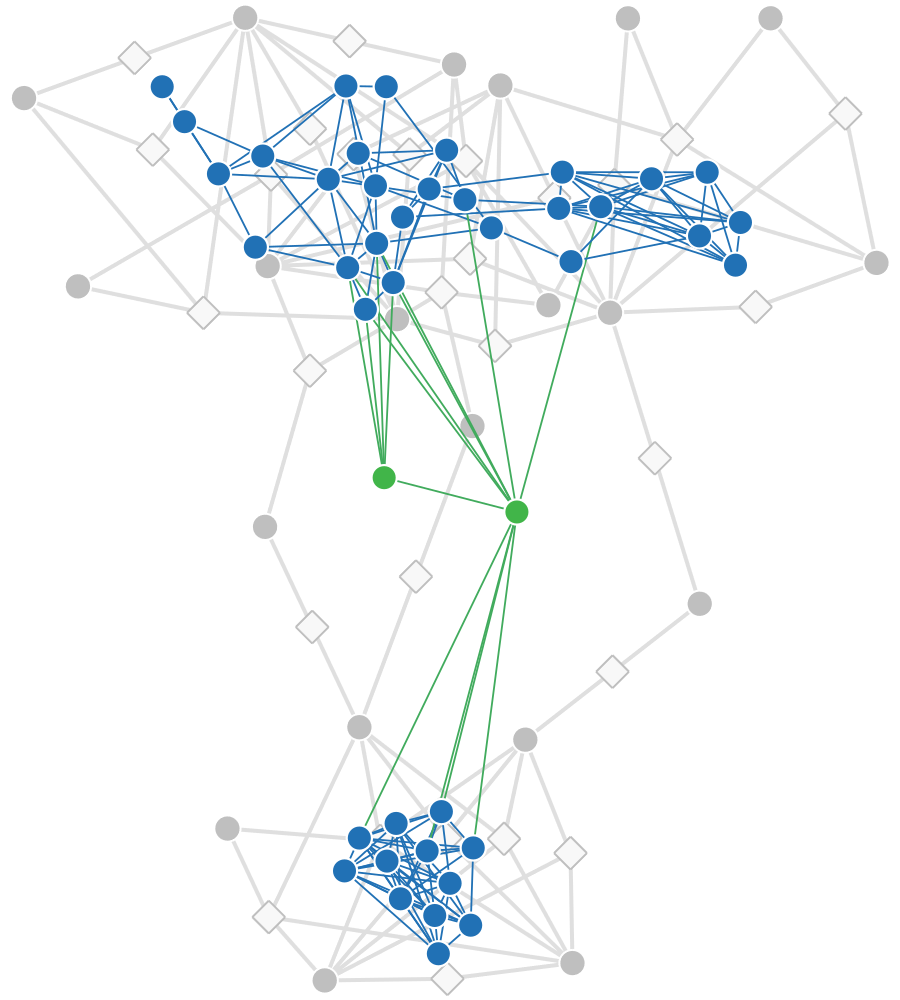}} \hspace{0.25in}
  \subfloat[][Clusters in primal]{\includegraphics[height=1.5in]{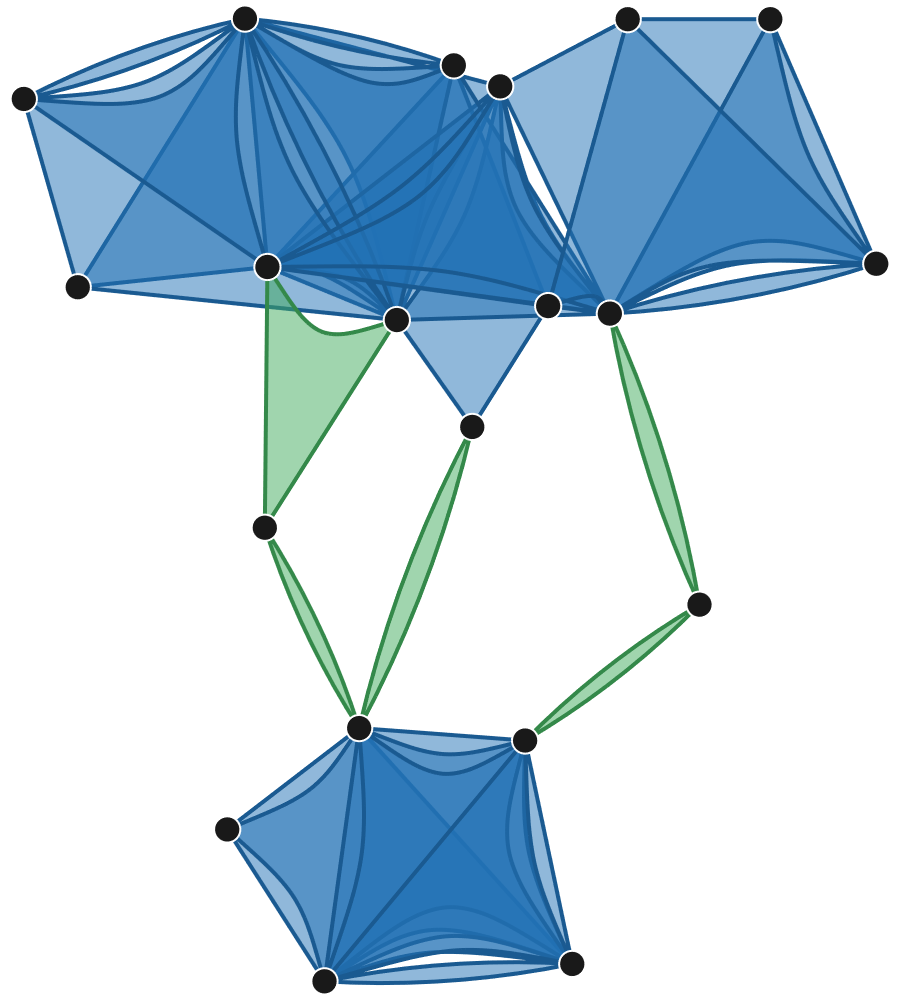}}
  \caption{An example of a topological block (a) containing multiple forbidden clusters. In (c), the cycle adjacency graph is superimposed over the bipartite representation (b), with the minimal basis cycles drawn as blue nodes, and the long basis cycles drawn as green nodes. Removing the green cycles leaves two 2-connected components of blue nodes, corresponding to the forbidden clusters highlighted in blue in (d).}
  \label{fig:clusters}
\end{figure*}

\subsection{Forbidden Sub-Hypergraphs}

A hypergraph $\HG$ is called \emph{Zykov planar} if its bipartite representation $G(\HG)$ is a planar graph~\cite{zykov1974hypergraphs}. A graph is planar, i.e. an edge crossing-free plane embedding can be found, if and only if it does not contain a subdivision of the complete graph $K_5$ or complete bipartite graph $K_{3,3}$~\cite{Kuratowski1930}. Zykov's definition for planarity assumes that hyperedges are drawn as arbitrary closed regions. However, Oliver \etal~\cite{oliver2024scalable} find this definition to be insufficient when drawing hyperedges as convex polygons, leading to a new definition for planarity within the polygon visualization metaphor:

\begin{definition}[Oliver \etal~\cite{oliver2024scalable}]
  A \textbf{convex polygon representation} is a drawing of a hypergraph in the plane where each hyperedge is represented as a strictly convex polygon such that the area of intersection between each pair of polygons is zero.
\end{definition}

A hypergraph that admits a convex polygon representation is called \emph{convex polygon planar}. Oliver \etal~\cite{oliver2024scalable} identify four \textit{forbidden sub-hypergraphs} that are Zykov planar but lack a convex polygon representation. They define an \emph{n-adjacent cluster} as the partial hypergraph induced by hyperedges $J \subseteq \E$ containing a common set of vertices $X \subseteq \V$ where $|X| = n \geq 2$. That is, set $X$ is contained within each hyperedge $e \in J$. To avoid confusion with other cluster definitions, we refer to this as an \emph{$n$-adjacent bundle of hyperedges}. The first forbidden sub-hypergraphs are a 3-adjacent bundle of 2 hyperedges (\Cref{fig:forbidden} (a1)) and a 2-adjacent bundle of 3 hyperedges (\Cref{fig:forbidden} (a2)). These two cases create a primal-dual pair. The next two forbidden sub-hypergraphs involve a \emph{strangled vertex} or \emph{strangled hyperedge} whose set of incident and adjacent elements have a particular structure. Oliver \etal~\cite{oliver2024scalable} describe a variant where a proper subset of the incident and adjacent elements form a cycle of length $n \geq 3$ (\Cref{fig:forbidden} (b1,b2)). We introduce another variant where a subset of the incident and adjacent elements form a star structure around a central element with $n \geq 3$ points (\Cref{fig:forbidden} (c1,c2)). The strangled vertex and strangled hyperedge cases also create primal-dual pairs.

\begin{theorem}[Oliver \etal~\cite{oliver2024scalable}] \label{thm:forbidden}
  Let $H$ be a Zykov planar hypergraph. Then $H$ has a convex polygon representation if and only if it does not contain any of the following as a sub-hypergraph:
  \begin{enumerate}[nosep]
    \item[(a)] A 3-adjacent bundle of 2 hyperedges,
    \item[(b)] A 2-adjacent bundle of 3 hyperedges,
    \item[(c)] A strangled vertex,
    \item[(d)] A strangled hyperedge.
  \end{enumerate}
\end{theorem}

In a polygon drawing of $T$, we observe that unavoidable overlaps have two causes: the forbidden sub-hypergraphs, which occur inside a local neighborhood of mutually incident and adjacent elements, and subdivisions of the Kuratowski subgraphs $K_5$ and $K_{3,3}$ in the bipartite representation which can occur among multiple long cycles. The $K_5$ and $K_{3,3}$ subdivisions can also occur inside a small radius of elements, but such instances often contain a forbidden sub-hypergraph within them. The forbidden sub-hypergraphs, on the other hand, are local structures by definition. We use \emph{local entanglement} and \emph{global entanglement} to describe these two categories of unavoidable overlap.

\subsection{Forbidden Clusters}

Let $\C$ be a cycle basis for the topological block $T$ containing only tight cycles, and let $G_T$ be the bipartite representation of $T$. We define the \emph{cycle adjacency graph} $A(\C)$ as a graph containing vertices for each basis cycle $C\in\C$ and edges $(C_i,C_j)$ for every bipartite edge in $G_T$ which the cycles $C_i$ and $C_j$ have in common. Note that $A(\C)$ is a non-simple graph since a pair of basis cycles can share multiple bipartite edges. The cycle adjacency graphs for each forbidden sub-hypergraph are shown in \Cref{fig:forbidden}, and an example for a larger topological block is shown in \Cref{fig:clusters}. Oliver \etal~\cite{oliver2024scalable} observe that the presence of forbidden sub-hypergraphs is correlated with the size of the intersection between pairs of adjacent vertices and pairs of adjacent hyperedges. We notice that every pair of shared vertices between hyperedges $e,f\in\E$ corresponds to a minimal cycle. This motivated us to study the subgraph of the cycle adjacency graph induced by the set of minimal basis cycles which we denote as $A(\C_4)$.

\begin{theorem} \label{thm:minimal}
  A cycle in $A_{(\C4)}$ defined by the sequence \mbox{$F=\langle C_1,C_2,\dots,C_k \rangle \subseteq\C_4$} for some tight cycle basis $\C$ of $T$ contains a common primal or dual node $x\in V(G_T)$ within each of the basis cycles $C_i\in F$ iff $F$ corresponds to a forbidden sub-hypergraph in $\HG$.
\end{theorem}

We prove \Cref{thm:minimal} in our supplementary material \Cref{apx:proof}. \Cref{thm:minimal} implies that forbidden sub-hypergraphs appear as cycles within $A(\C_4)$. We define a \emph{forbidden cluster} inside a topological block $T$ as a 2-connected component of the elements in $A(\C_4)$. This is a stronger notion than our definition of a topological block which includes any 1-connected component of elements in $A(\C)$. \Cref{fig:clusters} shows multiple forbidden clusters extracted from a single topological block. Under this definition, any forbidden sub-hypergraphs in $T$ must occur within a forbidden cluster. To extract the forbidden clusters of $T$, we re-use Hopcroft and Tarjan's algorithm to detect blocks of elements in $A(\C_4)$. Each of the non-trivial blocks extracted by this algorithm constitutes a new forbidden cluster in $T$.

\section{Structure-Aware Simplification}
\label{sec:simplification}

To achieve a meaningful multi-scale representation of $\HG$, we present a simplification method using operations specifically designed for the individual structures in our topological decomposition. We use three atomic simplifications that operate on the bipartite representation $G$: a minimal cycle collapse operation, a cycle edge cut operation, and a leaf pruning operation. The goal of our simplification is first, to eliminate non-planarity caused by local and global entanglement, and second, to reduce the space required by each structure in a polygon layout to facilitate high quality visualizations. A major difference in Oliver~\textit{et al.}'s~\cite{oliver2024scalable} approach is that they initialize candidate operations for every vertex and hyperedge in $\HG$, while we use the topological decomposition $D$ to generate candidate operations only where they are needed.

Another difference is the explicit use of both topology preserving and topology altering simplification operations. Oliver \etal use element removal and merger operations, both of which have the potential to maintain or alter the topology of $\HG$ depending on the configuration of the operand elements. For example, a series of merger operations could be used to collapse a cycle into a single vertex or hyperedge, reducing the first Betti number $B_1$ by one. On the other hand, a series of merger operations on a bridge or branch structure would not affect $B_1$ or the number of connected components $B_0$. In this paper, we use operations that are either always topology preserving, or always topology altering, allowing us to simplify $\HG$ in a more controlled way.

The topological decomposition $D$ of $\HG$ is an edge-disjoint union of topological blocks $T_i$, bridges $K_j$, and branches $K_k$. The topological blocks are defined by a tight cycle basis $\C=\{C_1,\dots,C_p\}$. While each of the structures in $D$ can be simplified in parallel, our implementation simplifies each structure sequentially in order of decreasing entanglement index. As a result, the bridges and branches, which contain no entanglement are simplified last. Our reasoning is based on experimental observations that non-planarity, which only occurs within the topological blocks, is the primary source of visual clutter in polygon visualizations. Of course, the sources of visual clutter can be data-dependent, and our simplification framework allows for any measure to be used in place of the entanglement index.

\begin{figure*}[btp]
  \centering
  \includegraphics[width=1.625in]{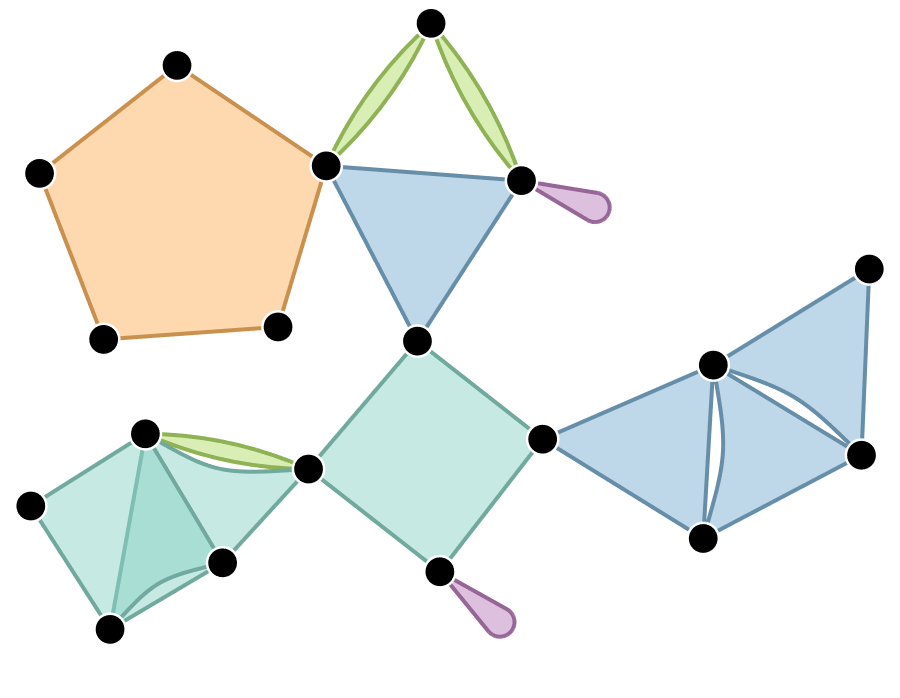} \hspace{0.125in}
  \includegraphics[width=1.625in]{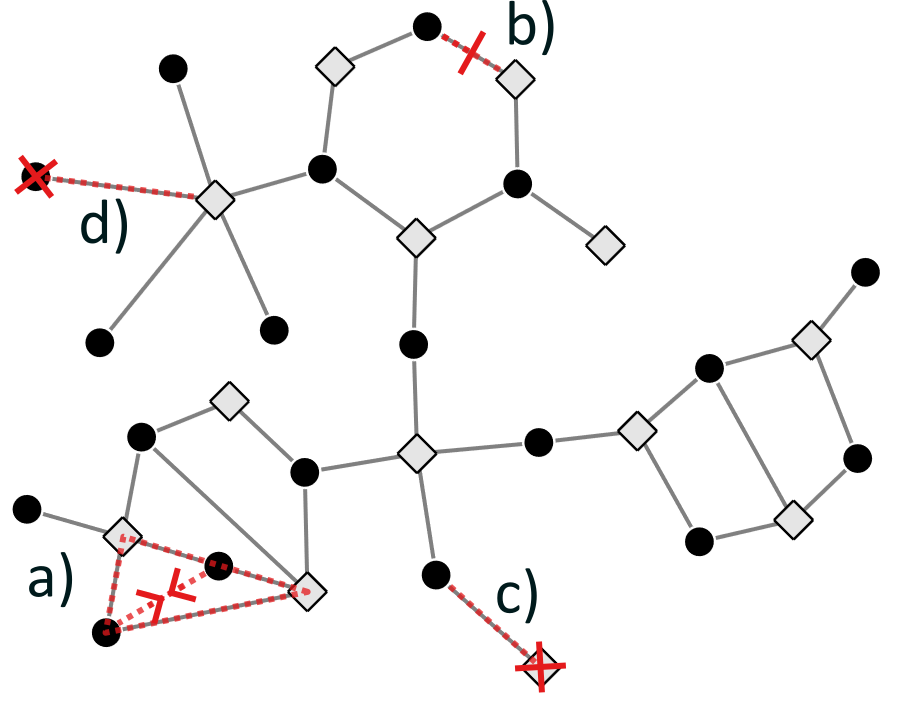} \hspace{0.125in}
  \includegraphics[width=1.625in]{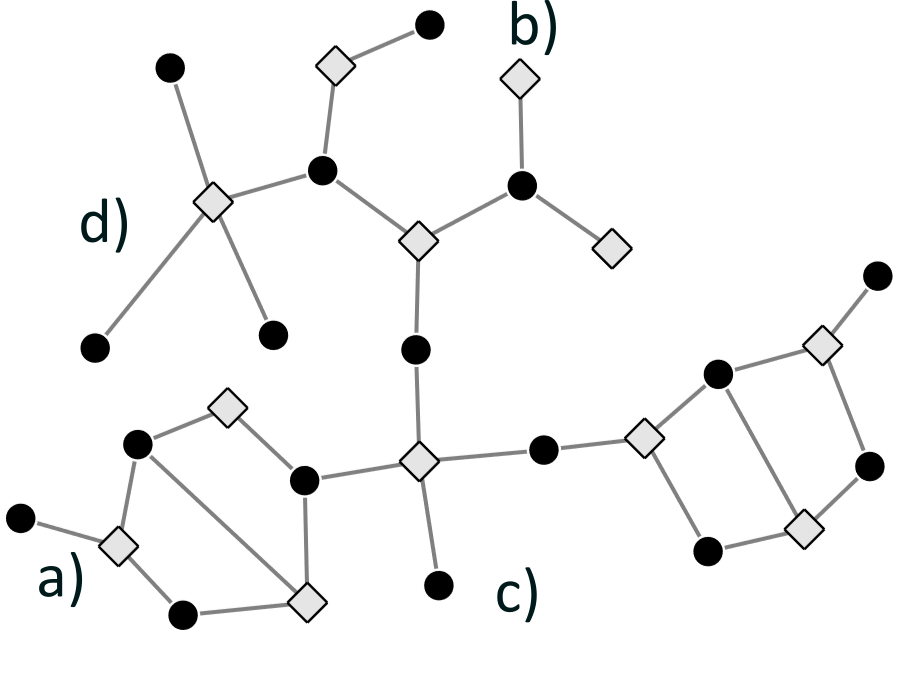} \hspace{0.125in}
  \includegraphics[width=1.625in]{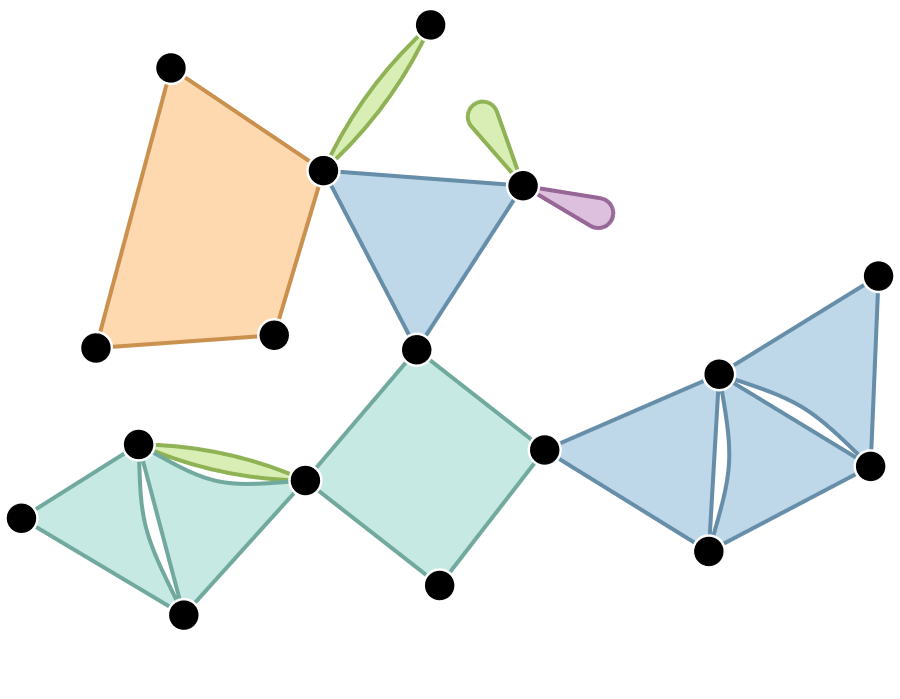}
  \caption{Example simplification operations in the hypergraph from \Cref{fig:bipartite,fig:topo_decomp}. On the far left and far right, we have the primal hypergraph before and after simplification. In the middle figures, we have four labeled operations (a), (b), (c), and (d). Operation (a) is a minimal cycle collapse, operation (b) is a cycle edge cut, operation (c) is a leaf pruning on a bridge, and operation (d) is a leaf pruning on a branch.}
  \label{fig:operations}
\end{figure*}

\subsection{Topological Blocks}

A topological block $T$ can contain both local entanglement in the form of forbidden sub-hypergraphs, and global entanglement in the form of Kuratowski subgraphs. We design a pair of topology altering operations to specifically target and reduce both types of entanglement, thereby reducing the amount of unavoidable overlaps in $T$.

The first operation we call a \emph{minimal cycle collapse}. Let $C=\langle u,e,v,f \rangle$ be a minimal cycle in $T$ where $u,v\in V(G)$ are primal nodes and $e,f \in E(G)$ are dual nodes. Collapsing $C$ into a single node has the potential to make $G$ non-bipartite and the hypergraph $\HG$ invalid. Instead, we collapse $C$ by merging together either the primal nodes $u,v$ or the dual nodes $e,f$. \Cref{fig:operations} (a) shows an example of a minimal cycle collapse using primal nodes. This operation alters the topology of $\HG$ by removing one or more minimal cycles depending on the direction of merging, reducing the value of $B_1$ accordingly. If $C$ is collapsed by merging primal nodes $v,u$, any of the other minimal cycle $C_i$ containing $u$ and $v$ are also collapsed and are removed from the cycle basis $\C$. A longer cycle $C_j$ containing $u$ and $v$, where $|C_j|>4$, is not at risk of being collapsed, but its length is reduced by 2.

The purpose of the minimal cycle collapse operation is to eliminate forbidden sub-hypergraphs. To identify candidate collapse operations, we first extract the set of forbidden clusters from $T$. If $T$ does not contain any forbidden clusters, it cannot contain any forbidden sub-hypergraphs, and we do not need any cycle collapse operations. Otherwise, for each forbidden cluster in $T$, we search for tight cycles within $A(\C_4)$, reusing our tight cycle basis algorithm from \Cref{sec:algorithm}. Let $F = \langle C_1, \dots, C_n \rangle$ be a tight cycle in $A(\C_4)$. If the minimal cycles $C_i\in F$ share any common bipartite nodes, we save $F$ as a forbidden sub-hypergraph and classify it according to the number and types of the shared bipartite nodes, as in the proof of \Cref{thm:minimal}. If $F$ corresponds to a forbidden sub-hypergraph, we identify a candidate cycle collapse operation for every $C_i\in F$.

The second operation we call a \emph{cycle edge cut}. Given a basis cycle $C_i$, we cut one of the edges $(v,e)\in E(C_i)$ in the bipartite graph, breaking the cycle $C_i$, and removing the connection between the corresponding vertex and hyperedge in $\HG$, as shown in \Cref{fig:operations} (b). This operation also alters the topology of $\HG$ by breaking one of the basis cycles, reducing the value of $B_1$ accordingly. Any basis cycles $C_j\neq C_i$ containing the edge $(v,e)$ are also affected and must be updated by replacing $(v,e)$ with the new shortest path between $v$ and $e$.

The purpose of the cycle edge cut operation is to eliminate instances of $K_5$ and $K_{3,3}$ occurring between non-minimal basis cycles. Thus, to identify edge cut candidates, we first construct a modified copy $T'$ by replacing each forbidden cluster in $T$ with a single node, retaining the external connections of the cluster. We note that $T'$ may no longer be a bipartite graph since the forbidden clusters contain both primal and dual nodes. For this reason, we only use $B'$ to search for instances of $K_5$ and $K_{3,3}$ and not as a representation of the original hypergraph. We then build an embedding of $T'$ with minimized edge crossings using a subgraph planarization algorithm from the Open Graph Drawing Framework~\cite{chimani2013ogdf}. We repeat the edge insertion phase of the planarization algorithm with ten permutations of the edge insertion order, taking only the best result with the fewest edge crossings. For each crossing between edges $(v_i,e_i),(v_j,e_j)\in E(T')$, we identify a candidate cycle edge cut operation for both $(v_i,e_i)$ and $(v_j,e_j)$.

Once all of the candidate minimal cycle collapse and cycle edge cut operations for topological block $T$ have been identified, we apply the priority ranking system of Oliver \etal~\cite{oliver2024scalable} to guide the order of application for each operation. Their system uses a priority measure containing multiple terms for per-element statistics including degree and cardinality, and betweenness centrality, each controlled by tuning parameters $\alpha,\beta$ and $\gamma$. We add to this an additional term and tuning parameter $\delta$ to evaluate how much each operation alters the topology of the input hypergraph $H$.

Let $O(x_1,x_2)$ be a candidate minimal cycle collapse or cycle edge cut operation. In the case that $O$ is a cycle collapse operation, let $x_1$ and $x_2$ represent the pair of primal or dual nodes to be merged in the collapse of minimal basis cycle $C$. In the case that $O$ is a cycle cut operation, let $x_1$ be the primal type node and $x_2$ the dual type node of the edge $(x_1,x_2) \in T$ to be cut. Let $s$ be the potential change in the first Betti number $B_0$ caused by the application of $O$. If $O$ is a minimal cycle collapse, $s$ is the number of minimal basis cycles in $\C$ containing $x_1$ and $x_2$. if $O$ is a cycle edge cut, $s$ is the total number of basis cycles in $\C$ containing the edge $(x_1,x_2)$. Now let $l$ be the average length of all basis cycles containing $x_1$ and $x_2$. We use $\delta \left(\frac{1}{s} + \frac{1}{l}\right)$ as the additional term in our operation priority measure. This term gives higher priority to operations that eliminate the fewest cycles, and lower priority to operations that affect long cycles. Our reasoning for prioritizing operations that eliminate the fewest cycles is straightforward, we aim to alter the topology of $\HG$ as little as possible. We prioritize preserving long basis cycles over short basis cycles for two reasons. Firstly, the long basis cycles can be considered more topologically significant while shorter basis cycles can be considered topological noise. Secondly, the long basis cycles are less likely to constrict the layout in a polygon drawing of $\HG$, and tend to pose fewer challenges for layout optimization algorithms.

After an operation $O$ has been applied, we update any remaining candidate operations that may have been affected. If $O$ is a cycle edge cut operation, we find the other edge in the same crossing identified by the planarization algorithm and remove it from consideration since the crossing has been resolved. If $O$ is a cycle collapse operation on a minimal cycle $C_i$ belonging to some forbidden sub-hypergraph $F_i$, we remove from consideration any collapse operations on the other minimal cycles $C_j\in F$, unless $C_j$ also belongs to another forbidden sub-hypergraph $F_j$ that has not yet been simplified. The simplification of $T$ terminates whenever the entanglement index $\eta(T)$ drops below a user defined threshold, or when $T$ is free from both forbidden sub-hypergraphs and subdivisions of the Kuratowski subgraphs, \ie when $T$ has a convex planar representation.

\begin{table*}[!t]
  \caption{Comparison of hypergraph simplification methods for reducing forbidden sub-hypergraphs in paper-author datasets. Each dataset is collected from the DBLP database~\cite{ley2009dblp} and consists of a maximal connected subset of publications in the specified year range from the IEEE journals Transactions on Pattern Analysis and Machine Intelligence (TPAMI) and Transactions on Visualization and Computer Graphics (TVCG). Each simplification was run until the input no longer contained forbidden sub-hypergraphs. Note that our method is more efficient and uses fewer operations to achieve this.}
  \vspace{-8pt}
  \begin{threeparttable}
    \scriptsize
    \setlength{\tabcolsep}{4.7pt}
    \centering
    \begin{tabular}{c c c c c p{0.0in} c >{\columncolor[rgb]{0.9,0.9,0.9}}c >{\columncolor[rgb]{0.8,0.8,0.8}}c p{0.0in} c >{\columncolor[rgb]{0.9,0.9,0.9}}c >{\columncolor[rgb]{0.8,0.8,0.8}}c}
      \toprule
      \multicolumn{5}{c}{\textbf{Dataset}} & & \multicolumn{3}{c}{\textbf{Oliver et al.}\cite{oliver2024scalable}} & & \multicolumn{3}{c}{\textbf{Ours}} \\
      \cmidrule{1-5}\cmidrule{7-9}\cmidrule{11-13}
      description & $|\V|$ & $|\E|$ & $B_1$ & $\eta(\HG)$ & & num ops. & initialization (s) & simplification (s) & & num ops. & decomp. (s) & simplification (s) \\
      \hline
        \rule{0pt}{1.25EM}
      TVCG (2015-2017) & 1008 & 429 & 334 & 0.234 &            & 275 & 0.305 & 0.556 &     & 83 & 0.012 & 0.006  \\
      TPAMI (2013-2020) & 2054 & 947 & 805 & 0.268 &           & 581 & 1.298 & 3.902 &     & 180 & 0.045 & 0.025 \\
      TVCG (2013-2020) & 3460 & 1570 & 1898 & 0.346 &          & 1452 & 4.190 & 845.0 &    & 429 & 0.101 & 0.094 \\
      \bottomrule
    \end{tabular}
  \end{threeparttable}
  \label{tab:comparison}
\end{table*}

\subsection{Bridges and Branches}

Let $K$ be a bridge or branch structure with $r\geq 1$ roots. Since $K$ does not contain any entanglement, our primary goal is to facilitate high-quality polygon layouts by reducing the amount of space required by the individual hypergraph elements in $K$. We do this using a \emph{leaf node pruning} operation. This is similar to the cycle edge cut operation in that it breaks the incidence relationship between a vertex $v_i\in\V(K)$ and a hyperedge $e_i\in\E(K)$. However, the leaf pruning operation requires that either $v_i$ or $e_i$ correspond to a leaf node in $G$, and deletes the leaf node after cutting the bipartite edge $(v_i,e_i)$. Furthermore, the leaf pruning operation is topology preserving while the edge cut operation is topology altering. A leaf node cannot belong to a cycle, so pruning does not affect $B_1$. Since the leaf node is deleted after the edge is cut, pruning also has no impact on the number of connected components $B_0$. An example of pruning a dual leaf node from a bridge is shown by \Cref{fig:operations} (c), and an example of pruning a primal leaf node from a branch structure is shown by \Cref{fig:operations} (d).

We identify leaf nodes in $K$ using a multi-source breadth-first search starting from the root nodes of $K$. For each leaf node discovered, we identify a new pruning operation candidate. Within the search, we also keep track of the depth $d(x)$ of each node $x$ with respect to the roots of $K$ as well as the furthest depth $d_{low}(x)$ reached by any of the children of $x$. A root node $r$ of $K$ has $d(r)=0$ and $d_{low}(r)$ equal to the total height of the tree while a leaf node $x$ has $d_{low}(x)=d(x)$.

Once the candidate pruning operations have been identified, we again use the operation priority ranking system of Oliver \etal to determine the order of simplifications. For bridge and branch structures we add the term $\delta\left(1-\frac{d_{low}(x)}{d_{low}(r)}\right)$ to the operation priority measure where $x$ is the leaf node in a candidate operation and $r$ is a root node of $K$. This term promotes preserving the deepest elements in $K$ as well as the elements on the connecting paths between $r$ and the deepest elements. We use this to help preserve the longest paths in each bridge and branch which allows them to be easily identified in visualizations of simplified scales.

After pruning a leaf node $x$, we check whether the parent of $y$ of $x$ has become a leaf node. If so, we identify a new candidate pruning operation for $y$ and add it to the priority list of candidate operations. The simplification terminates once the priority of the next pruning operation drops below a user-defined threshold.

\section{Results}
\label{sec:results}

To evaluate the performance of our structure-guided simplification, we compare it to the priority guided method of Oliver \etal~\cite{oliver2024scalable}. They take a statistical approach where iterative atomic simplifications are guided by a priority measure containing terms for the distributions of different per-element statistics. In particular, they use terms for the distribution of vertex degrees and hyperedge cardinalities, vertex and hyperedge adjacency factors, and betweenness centrality. The influence of each distribution on the order of simplifications is controlled by tuning parameters $\alpha,\beta,\gamma$. The purpose of their adjacency factor term, which measures the volume of shared vertices between a hyperedge and all of its adjacent hyperedges, is to guide the simplification toward eliminating forbidden sub-hypergraphs.

We compare the efficiency of both approaches in \Cref{tab:comparison} for the task of removing forbidden sub-hypergraphs from three different hypergraph datasets. For each method, we list the number of operations required to remove all forbidden sub-hypergraphs from the input, the time required to initialize the operations, or in our case, perform the decomposition, and the time required to apply the operations and perform any necessary updates. We observe that both approaches increase in the number of operations and time required as the datasets increase in size as well as entanglement index $\eta(\HG)$. However, our approach requires significantly fewer operations and significantly less execution time. The smaller number of operations can be attributed to our decomposition and extraction of forbidden clusters which tell us exactly where the forbidden sub-hypergraphs live. The difference in execution time, especially for the largest dataset, is even greater. This can be due to the fact that we simplify each decomposition structure independently, maintaining separate operation priority queues for the operations on each structure, while Oliver \etal maintain a much larger global priority queue that needs to be updated and potentially re-sorted after every atomic simplification. Finally, we observe that their implementation is iterative in nature relying on an updating scheme that does not have an obvious parallel implementation. Ours has the potential to be sped up even further by simplifying the decomposition structures in parallel.

\begin{figure*}[btp]
  \raisebox{0.75in}{\includegraphics[height=1.25in]{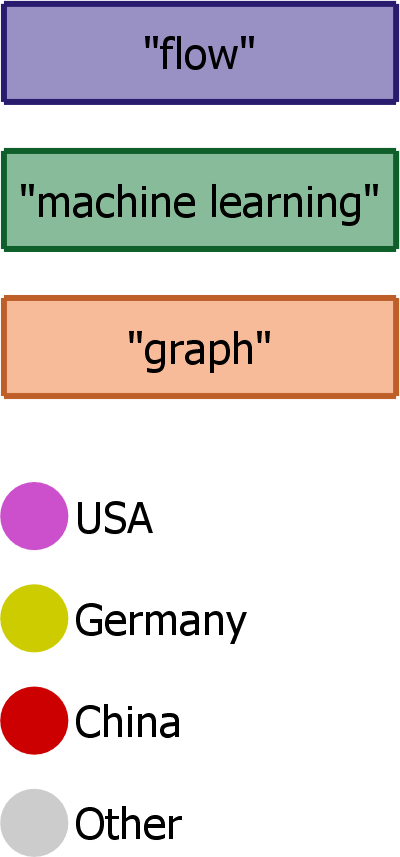}}
  \subfloat[][Input hypergraph]{\includegraphics[height=2.0in]{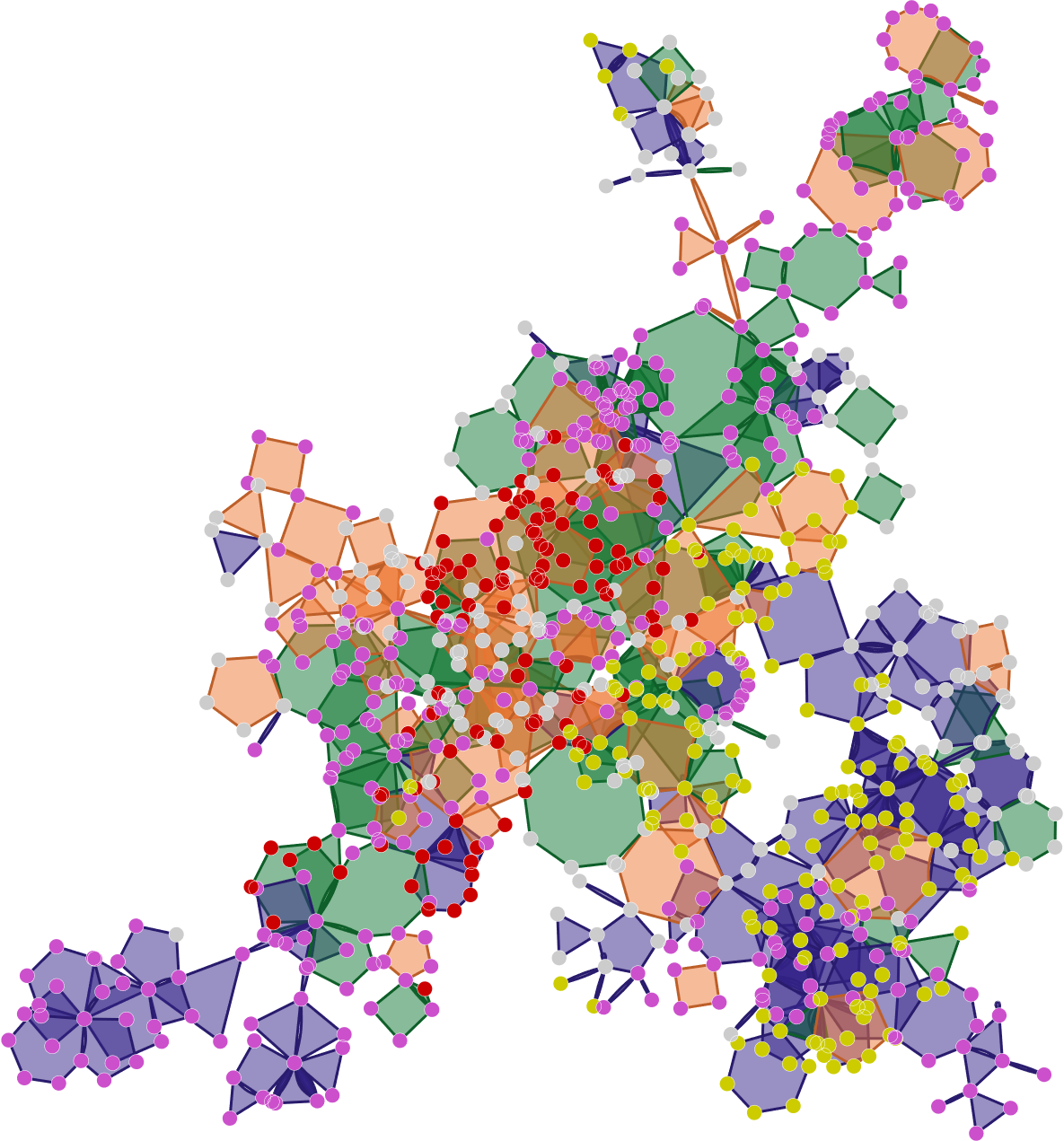}} \hspace{0.25in}
  \subfloat[][Oliver \etal simplified scale]{\includegraphics[height=2.0in]{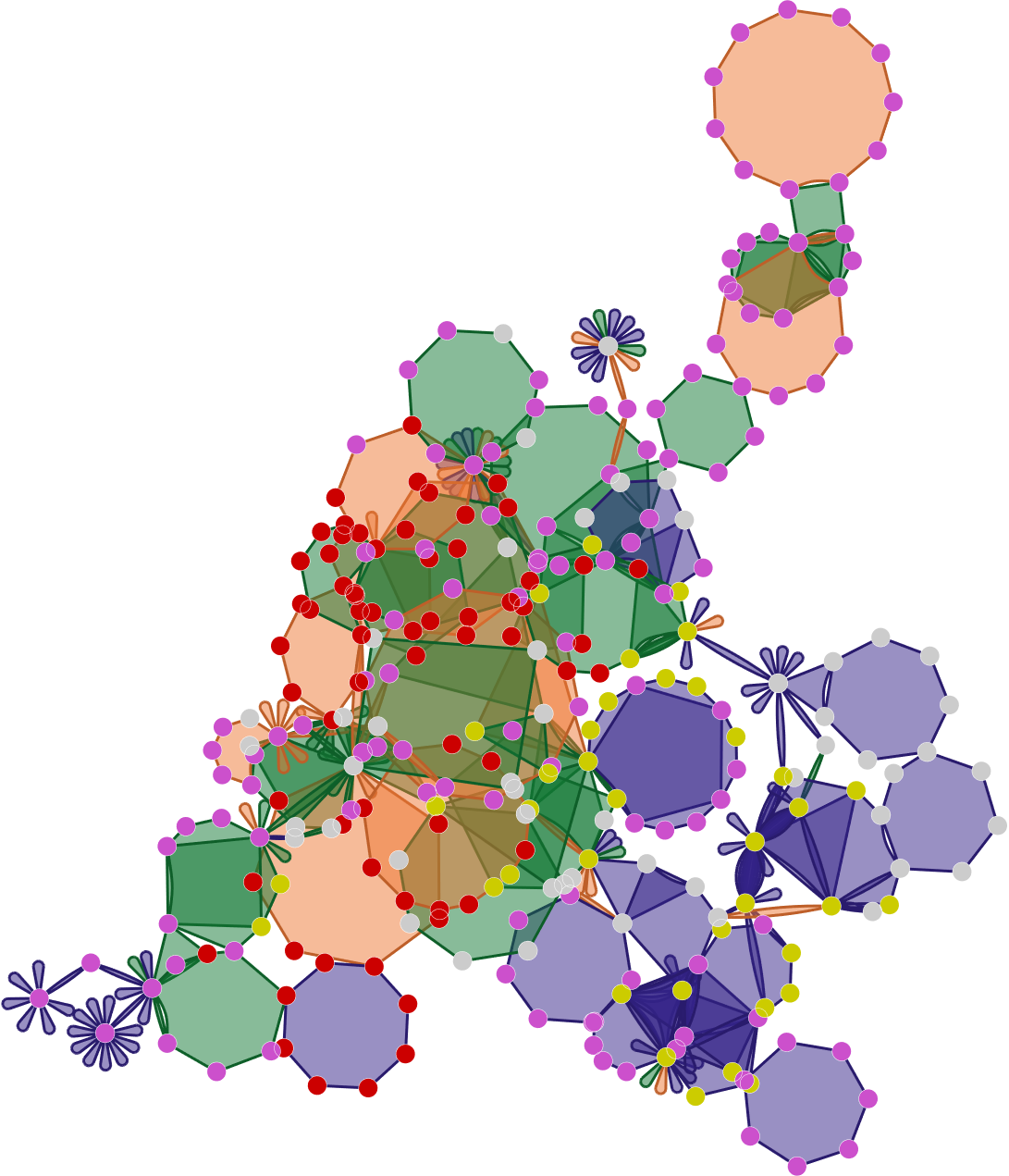}} \hspace{0.25in}
  \subfloat[][Our simplified scale]{\includegraphics[height=2.0in]{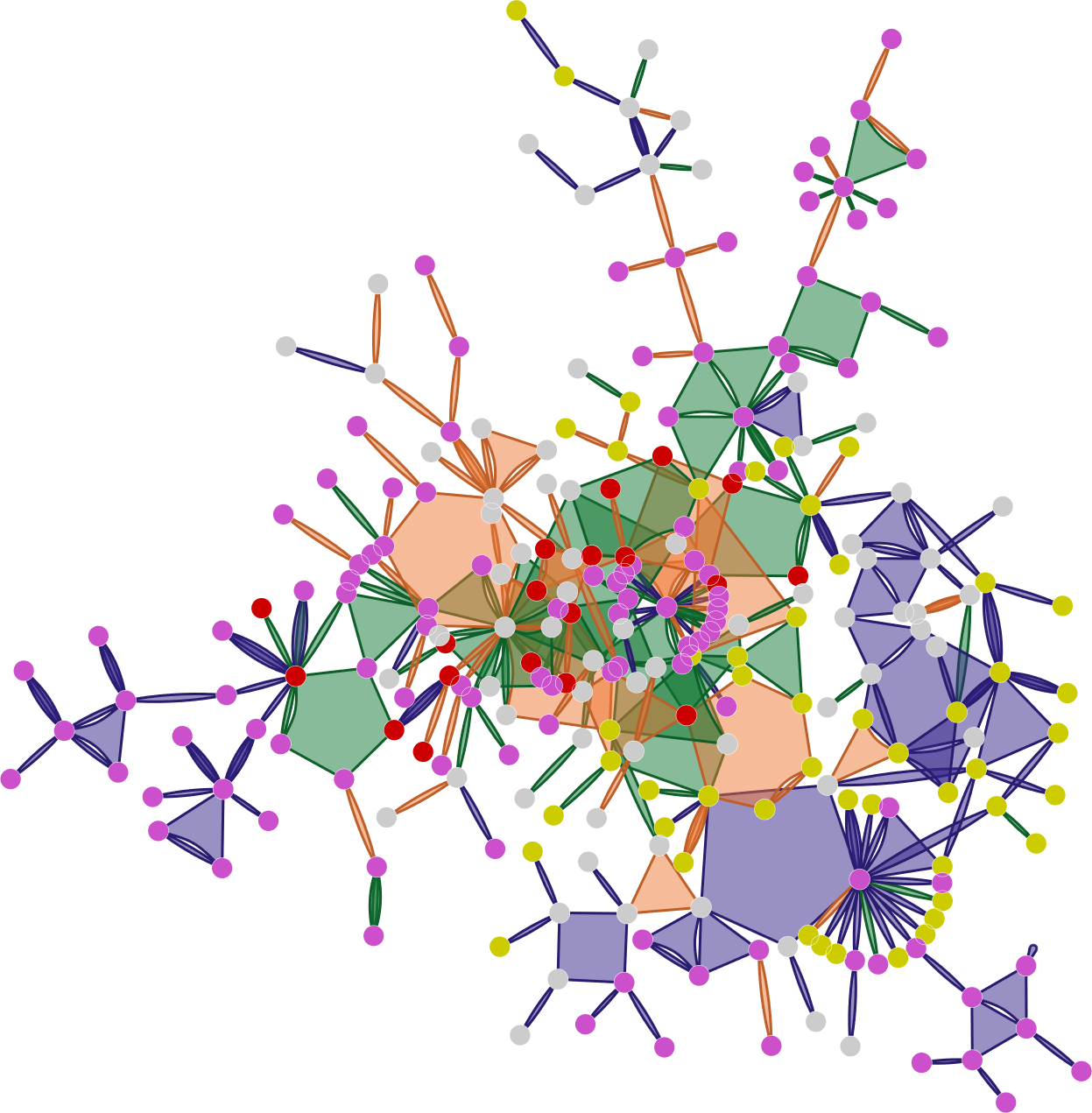}} 
  \caption{A paper-author hypergraph network with 786 vertices and 318 hyperedges, (a) is simplified using the priority guided approach of Oliver \etal~\cite{oliver2024scalable} (b), and our topological decomposition guided approach (c). Notice that some of the branches in (b) have been reduced and others have been eliminated entirely. Our hypergraph decomposition extracts structures ahead of time, allowing us to preserve the skeleton of each branch. In each visualization, the hyperedges are colored according to the keywords of their papers: blue for the keyword ``flow'', green for the keywords ``machine learning'', and orange for the keyword ``graph''. Additionally, the vertices are colored according to the geographic location of the affiliated research institution for each author: magenta for institutions based in the United States, yellow for institutions based in Germany, red for institutions based in China, and gray for institutions based elsewhere. Enlarged versions of these visualizations are provided in our supplementary material \Cref{apx:figures}.}
  \label{fig:paperauthor}
\end{figure*}

To evaluate the utility of our decomposition and simplification framework, we apply it to three real-world datasets. For a simplified result $\HG'$, we generate high quality polygon layouts by first applying a force directed layout to the simplified bipartite graph $G'$. We align the vertices of $\HG'$ with the corresponding primal nodes in the layout of $G'$, and draw the hyperedges as polygons between their contained vertices using the starrization technique of Qu \etal~\cite{Qu:22}. We then apply the automatic primal-dual polygon layout optimization system of Qu \etal to $\HG'$. Their optimization uses a multi-term objective function that promotes polygon regularity and uniform side lengths while avoiding unnecessary polygon overlaps.

\vspace{0.0625in}
\noindent \boldpara{Friendship Dataset} This dataset involves friendships between high school students in Marseilles, France, recorded in 2013~\cite{mastrandrea2015contact}. The original dataset is a directed network of reported friendships among the students. We construct an undirected graph from this dataset including only edges where both students reported being friends with each other. We then construct a hypergraph $\HG$ by creating a hyperedge for every maximal clique in this graph. A connected subset of $\HG$ is visualized in \Cref{fig:teaser} (a). Each hyperedge represents a friendship group where every student reported being friends with every other student in the group.

The topology can tell something about the different communities in $\HG$. For example, in relation to the zeroth Betti number $B_0$, the connected components of $\HG$ suggest larger communities among the students, possibly representing the individual classes from the dataset described in \cite{mastrandrea2015contact}. In relation to the first Betti number $B_1$, the cycles within each topological block can tell us how \emph{close-knit} a particular community is. For instance, a community consisting of many small cycles is more close-knit than a community with only long cycles, which is in turn more close-knit than a community that has a tree structure. Thus, being able to identify the existence and length of cycles in $\HG$, related to the entanglement index of its topological blocks, can be a valuable tool for studying the different communities.

\Cref{fig:teaser} (a) highlights in blue a length 4 cycle of $\HG$ that occurs between two forbidden clusters. In (b), a simplified scale $\HG'$ is produced using the method of Oliver \etal~\cite{oliver2024scalable} with tuning parameters $\alpha=0.0,\beta=0.9,\gamma=0.4$. Notice that the vertices in the highlighted cycle have been collapsed into a single vertex, combining the forbidden clusters on either side. This gives the impression of a single close-knit group of students in $\HG'$ where there is actually some separation between the clusters in $\HG$. In (c), the highlighted cycle is preserved after applying our structure-aware simplification method, and the original forbidden clusters from $\HG$ remain visually distinct. Our simplification $\HG''$ was produced using the same number of simplification operations as $\HG'$, the same values for $\alpha,\beta,\gamma$, and a value of $\delta=1.0$ for our new priority terms. We also observe that the bottom half of $\HG'$ is nearly identical to $\HG$, and still contains a few forbidden sub-hypergraphs. Our result $\HG''$ on the other hand does not contain any forbidden sub-hypergraphs. This indicates that our decomposition-based approach is a more targeted and efficient way to reach a planar result.

The result in (c) was reached after applying 25 minimal cycle collapse operations and 1 cycle edge cut operation. This indicates that the original hypergraph contains more local entanglement in the form of forbidden clusters than global entanglement in the form of $K_5$ or $K_{3,3}$ subdivisions. These forbidden clusters can indicate broader friendship circles than the individual hyperedges would imply. While it would be possible to cut open these friendship circles to reach planarity, it seems more appropriate to eliminate the forbidden sub-hypergraphs by merging some of their elements. This way, some distinction between individual elements is lost, but the connections within the friendship circle are emphasized. For example, the overlapping red and orange polygons on the left side of (a) are merged into a single octagon in (c), filling in the few missing connections among the friendship circle. On the other hand, two vertices being merged together can be interpreted as a pair of students who have very similar friendship connections. This led to the appearance of \emph{monogons} (purple teardrop shapes) in (c) which indicate tighter friendship groups within a merged ``super-vertex''.

The cycle cut operation that is applied in (c) removes one friend from a friendship group. Our simplification system is designed such that the cut friendship occurs along a non-minimal cycle, and does not cut open a close-knit forbidden cluster. Furthermore, we prioritize cutting bipartite edges that occur along the fewest and shortest basis cycles. Thus, while the immediate connection is broken, the increase in the shortest path between the removed friend and their friendship group (along the other side of the broken cycle) is minimized. In this way, our simplification framework balances maintaining both path length and path structure by using both collapse and cut operations, but the acceptability of artifacts from one or the other may depend on the specific dataset and required analysis tasks.

\vspace{0.0625in}
\noindent \boldpara{Paper-Author Dataset} Isenberg et al.'s Vispubdata dataset~\cite{Isenberg:2017:VMC} contains publication information for IEEE Visualization papers from 1990-2021. \Cref{fig:paperauthor} (a) shows a connected subset of publications containing the keywords "flow", "graph", and "machine learning", where each author is a vertex and each paper a hyperedge. The vertices are colored according to the geographic location of each author's affiliated research institution: magenta for institutions based in the United States, yellow for institutions based in Germany, red for institutions based in China, and gray for institutions based elsewhere. The hyperedges are colored according to the keywords of the corresponding papers: blue for the keyword ``flow'', green for the keywords ``machine learning'', and orange for the keyword ``graph''. For papers containing multiple of these keywords, we choose the color based on which keyword appears first. This dataset contains numerous bridge and branch structures as well as several large cycles. Entangled topological blocks can be interpreted as topic areas with a high level of cross-collaboration within or between research groups. Based on the density of overlapping polygons in \Cref{fig:paperauthor} (a), there appears to be much cross-collaboration between researchers based in Germany on the topic of flow visualization and cross-collaboration between researchers based in China on machine learning and graph visualization topics. However, it is difficult to tell whether there are actually cycles present among the overlapping hyperedges or if this is just a result of the layout. Our simplified result in (c) was produced using only minimal cycle collapse and leaf pruning operations. In this layout, all of the forbidden sub-hypergraphs are eliminated, but we still see a number of cycles and overlapping hyperedges among the Germany and China based researchers. This confirms our observations of cross-collaboration among these research communities. The result in (b) was produced by running the simplification algorithm of \cite{oliver2024scalable} with the same tuning parameters until reaching a scale with the same number of elements as (c). Their method does not guarantee that large cycles are preserved, so we cannot reach the same conclusion based on their visualization result.

We also see in \Cref{fig:paperauthor} (a) that many of the US based researchers appear to be connected to the rest of the hypergraph through long bridge or branch structures. These structures can represent new research directions or new connections between existing topic areas. The large bridge and branch structures are even more apparent in (c), but we also discover many short branches that are hidden among the overlapping hyperedges in (a) and (b). This indicates that in addition to the cross-collaboration seen in the topological blocks, there are many papers connected to the rest of the community through a single author, who may be an advisor with numerous students, or an influential publication.

In \Cref{fig:paperauthor} (a), we can see two rabbit ear structures near the top extending from the central part of the hypergraph. In (c), we find that the rabbit ear structures include small topological blocks near the ends. In the context of our topological decomposition, the rabbit ears are the union of a bridge, topological block, and several small branches. The small topological blocks in each ear are important since they suggest a research group with frequent collaborations rather than an individual researcher with many one-time collaborators. In fact, the topological block in the left ear is formed by a group of three researchers, each pair of whom collaborated on a different topic. In (b), the left ear is reduced to a short branch with a high degree vertex at the end, giving the opposite impression of a sole prolific researcher. Furthermore, because the simplification method of Oliver \etal~\cite{oliver2024scalable} is iterative, the cycle in the left ear was collapsed in a previous simplified scale, changing the structure of the ear from a union of a bridge, cycle, and branches into a single branch. Thus, the simplified result in (b) involves both a false negative case in the collapsed cycle, and a false positive in the new branch structure replacing the original rabbit ear from (a).

The interpretation of minimal cycle collapse operations in this dataset is similar to the previous friendship dataset: papers or authors from the same close-knit research group are merged together to eliminate visual clutter inside forbidden clusters. The leaf nodes pruned in (c) could be considered less significant in terms of connectivity in the research community because they represent authors with only one paper in the field, possibly student research assistants, or papers with only one author. Iterative leaf pruning has the potential to eliminate a branch structure entirely, removing important connecting authors, which is why we prioritize preserving the deepest elements in each branch. However, for certain visualization tasks, such as comparing the number of authors on each paper or the number of publications for each author, leaf pruning may not be appropriate.

\begin{figure}[btp]
  \subfloat[][Input hypergraph]{\includegraphics[height=2.75in]{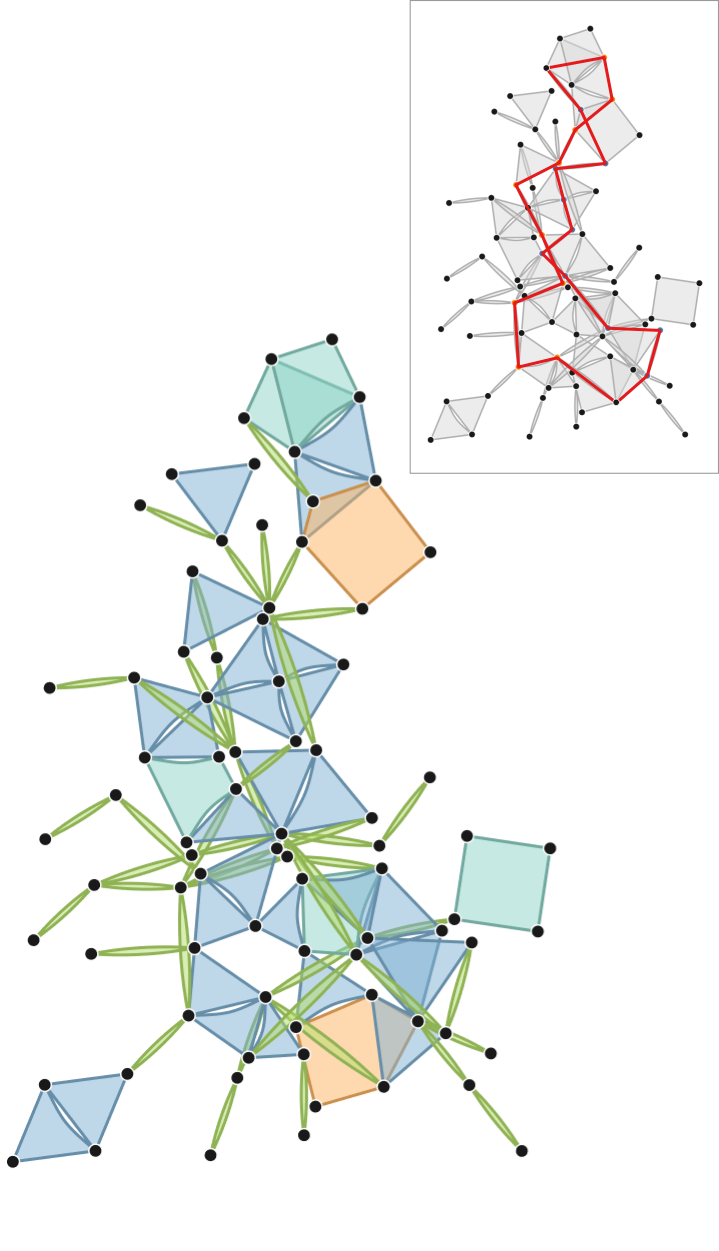}}
  \subfloat[][Simplified hypergraph with cycle edge cuts]{\includegraphics[height=2.75in]{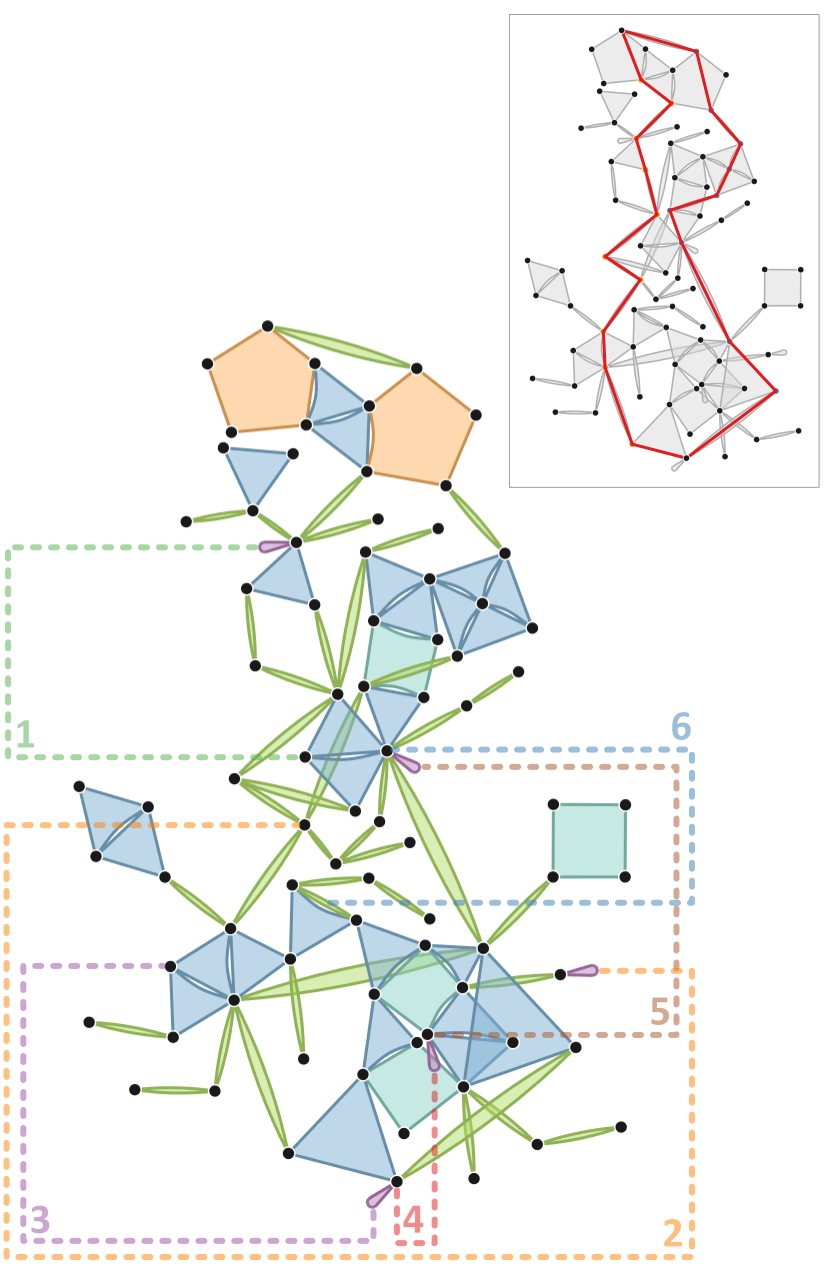}}
  \caption{A hypergraph representing face-to-face interactions between museum visitors \cite{isella2011s} is visualized before, (a), and after simplifying the topological blocks, (b). Notice the numerous triangles and digons in (a) that overlap despite being non-adjacent. On the right, six cycles have been cut to make the hypergraph planar. The deleted incidence relationships between vertices and hyperedges are represented by dashed annotation lines along the boundary of the visualization. In the upper right corners, we highlight the same two paths for each layout from a vertex near the top to a vertex near the bottom.}
  \label{fig:museum}
\end{figure}

\vspace{0.0625in}
\noindent \boldpara{Museum Interactions Dataset} Our final dataset from Isella \etal~\cite{isella2011s} tracks social contact patterns between visitors in a museum gallery. Face-to-face interactions between gallery visitors were measured using electronic badges. Qu \etal~\cite{Qu:22} construct hypergraphs from this data by creating hyperedges between maximal cliques of participants who spent more than 40 seconds interacting face-to-face with each other. \Cref{fig:museum} (a) shows the hypergraph $\HG$ for gallery visitors on May 5th, 2009. In (b), we show a planar simplified scale obtained using our topology altering simplification operations. We include dashed annotation lines to indicate the incidence relationships from the original hypergraph that are removed in the simplified scale. Such annotations in simplified results can provide a visual indication of the global entanglement present in the original hypergraph. We provide a comparison of our simplifications here to Zhou \etal~\cite{zhou2023simplification} in our supplementary material \Cref{apx:zhou_comparison}.

We observe that reaching a planar simplified scale $\HG'$ in this case requires 6 edge cycle cut operations and 2 minimal cycle collapse operations, meaning $\HG$ contains more subdivisions of the Kuratowski graphs than forbidden sub-hypergraphs. This indicates that although there are many overlapping polygons in the layout of $\HG$, the average adjacency between pairs of elements is relatively low. From \Cref{fig:museum} (a) we can already see that a majority of the face-to-face interactions only involve two or three people. The visualization in (b) further shows that there are not many common participants between separate interactions. This makes sense in a gallery setting where small groups of visitors who know each other have only a few interactions outside that group where they may be listening to a tour guide or speaking with a docent. 

In the context of tracking infectious diseases, the disentangled visualization in (b) gives a clearer view of the possible transmission vectors between visitors. The close proximity of several vertices in (a) combined with the overlapping polygons makes it difficult to tell how many distinct paths connect the uppermost elements in the layout to the bottommost elements. In the upper right corner of (a), two such distinct paths are highlighted in red, but there are several places in the layout where these paths appear to share a vertex. This could be improved by refining the layout, but the underlying issue remains that some amount of overlap is unavoidable since the hypergraph does not have a convex polygon representation. In (b), the vertices and hyperedges are more spread out, and it is easier to distinguish distinct paths between elements. In the upper right corner of (b), the same two distinct paths are highlighted in red, and we can clearly see that they do not intersect except at the uppermost and bottommost vertices.

\section{Conclusions and Future Work}
\label{sec:conclusion}

In this paper, we present a novel structure-based hypergraph decomposition using the topology of the bipartite graph representation. Within our decomposition, the use of the bipartite graph representation also allows us to identify entangled minimal cycles as the main source of unavoidable overlaps in hypergraph visualizations. This leads to a new definition of the entanglement index which is based on the ratio of the first Betti number and the number of vertices in a topological block. We use our decomposition to augment the atomic simplification framework of Oliver et al.~\cite{oliver2024scalable} making use of structure preserving and structure altering operations. We also provide efficient algorithms to compute the decomposition, as well as a framework for implementing structure-aware simplification. Compared with ~\cite{oliver2024scalable}, our work makes the interpretation of simplified hypergraph visualizations more reliable because the main structures in the data are preserved. An implementation of our decomposition, simplification, and visualization tools are available on GitHub: \url{https://github.com/peterdanieloliver/HGPolyVis}.

The goal of our work is to provide the theoretical groundwork for structure-based hypergraph simplification upon which future works can build. A thorough evaluation is needed to fully explore the potential visualization benefits, such as identifying disjoint paths in simplified hypergraphs. We consider our evaluations preliminary and plan to conduct additional user studies as future work.

We will collaborate with domain scientists in future work to assess the usefulness of our approaches and explore possible improvements for application dependent visualization problems. We also plan to more rigorously investigate the use of glyphs and annotations for communicating artifacts produced by our simplifications. While our work identifies entangled cycles as the source of unavoidable overlaps in polygon visualizations, requiring the polygons to be near-regular can also cause overlaps in the bridge and branch structures. We plan to investigate the use of hyperbolic spaces to address this issue. We also plan to explore structure-based hierarchical layout methods to make polygon visualization of hypergraphs with thousands to tens of thousands of hyperedges more tractable. Finally, we plan to study extensions of our work to time-varying hypergraphs as well as AR/VR.

\acknowledgments{
	The authors wish to thank the anonymous reviewers and paper chairs for their constructive feedback.
}

\bibliographystyle{abbrv-doi-hyperref}
\bibliography{hypergraph2024}

\clearpage
\appendix

\section{Cycle Basis Algorithm}
\label{apx:algorithm}

Here we present pseudocode for our tight cycle basis algorithm described in \Cref{sec:algorithm} of the main paper. The input is a topological block $T$ of some hypergraph $\HG$ and the output is a basis $\C$ for the cycle space of $T$ containing only tight cycles. Let $n=|V(T)|$ and let $m=|E(T)|$. The complexity of our tight cycle basis algorithm is $O(B_1(T)(n+m))$ where $B_1(T)$ is the first Betti number. The Euler characteristic gives us $B_1(T)=1+m-n$, so the complexity is in fact $O((m-n)(n+m))=O(m^2-n^2)$.

\begin{algorithm}
  \caption{BFS Tight Cycle Basis} \label{alg:basis}
  \KwIn{Topological block $T$}
  \KwOut{Tight cycle basis $\C$ of $T$}
  \SetKwBlock{Begin}{function}{end function}
  \Begin($\textsc{BfsBasis}{(}T{)}$)
  {
    $Q \gets$ initialize queue \tcp*{\textit{outer search queue}}
    $S \gets$ empty subgraph \tcp*{\textit{visited nodes and traversed edges}}
    choose node $x_0 \in V(T)$; \\
    enqueue $x_0 \rightarrow Q$; insert $x_0 \rightarrow V(S)$; \\
    \While(\tcp*[f]{\textit{outer BFS loop}}){$|Q|>0$}
    {
      $x \gets Q$.front; $Q$.pop; \\
      \For{each edge $(x,y)\in E(T),$ $(x,y)\notin E(S)$}
      {
        \If(\tcp*[f]{\textit{cycle detected}}){$y\in V(S)$}
        {
          $C =$ \textsc{TightCycle}$(S,x,y)$ \tcp*{\textit{call inner BFS}}
          add $C$ to homology basis $\C$;
        }
        \Else{
          enqueue $y\rightarrow Q$; insert $y\rightarrow V(S)$;
        }
        insert $(x,y)\rightarrow E(S)$;
      }
    }
    \Return{$\C$};
  }
  \Begin($\textsc{TightCycle}{(}S,x,y{)}$)
  {
    $Q \gets$ initialize queue \tcp*{\textit{inner search queue}}
    $U \gets$ empty subgraph \tcp*{\textit{visited nodes and traversed edges}}
    enqueue $x \rightarrow Q$; insert $x \rightarrow V(S)$; \\
    $x$.parent $\gets$ null; \\
    \While(\tcp*[f]{\textit{inner BFS loop}}){$|Q|>0$}
    {
      $a\gets Q$.front; $Q$.pop; \\
      \For{each edge $(a,b)\in E(S)$, $(a,b)\notin E(U)$}
      {
        \If(\tcp*[f]{\textit{tight cycle found}}){$b=y$}
        {
          $C\gets$ new cycle; insert $(a,y)\rightarrow C$; \\
          $p\gets a$.parent; \\
          \While{$p\neq$null}
          {
            insert $(p,a)\rightarrow C$; \\
            $a\gets p$; $p\gets p$.parent;
          }
          \Return{$C$};
        }
        \Else{
          enqueue $b\rightarrow Q$; insert $b\rightarrow V(U)$; \\
          $b$.parent $\gets a$;
        }
        insert $(a,b)\rightarrow E(U)$;
      }
    }
  }
\end{algorithm}

\section{Proof of \Cref{thm:minimal}}
\label{apx:proof}

\begin{figure}[tbp]
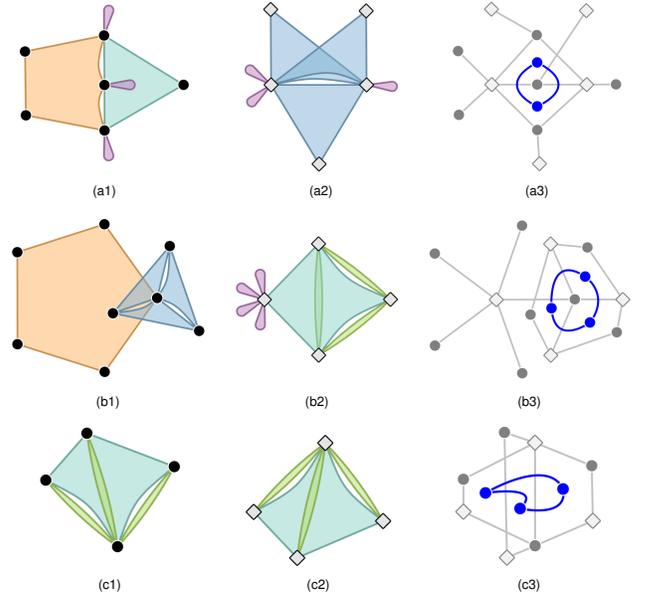

  \centering
  \captionsetup[subfigure]{labelformat=empty}
  \subfloat[][(a1)]{\includegraphics[height=0.9in]{3adjacent_primal.png}} \hspace{0.25in}
  \subfloat[][(a2)]{\includegraphics[height=0.9in]{3adjacent_dual.png}} \hspace{0.25in}
  \subfloat[][(a3)]{\includegraphics[height=0.9in]{3adjacent_adjacency.png}} \\
  \subfloat[][(b1)]{\includegraphics[height=0.9in]{strangled_primal.png}} \hspace{0.125in}
  \subfloat[][(b2)]{\includegraphics[height=0.9in]{strangled_dual.png}} \hspace{0.125in}
  \subfloat[][(b3)]{\includegraphics[height=0.9in]{strangled_adjacency.png}} \\
  \subfloat[][(c1)]{\includegraphics[height=0.75in]{saturated_primal.png}} \hspace{0.3in}
  \subfloat[][(c2)]{\includegraphics[height=0.75in]{saturated_dual.png}} \hspace{0.3in}
  \subfloat[][(c3)]{\includegraphics[height=0.75in]{saturated_adjacency.png}}
  \caption{The forbidden sub-hypergraphs of polygon hypergraph drawings: (a1) 3-adjacent hyperedge bundle of 2 hyperedges, (a2) 2-adjacent hyperedge bundle of 3 hyperedges, (b1) strangled vertex cycle variant, (b2) strangled hyperedge cycle variant, (c1,c2) strangled vertex and hyperedge star variant. Notice that (a2) is the dual of (a1), (b2) is the dual of (b1), and (c2) is the dual of (c1). The cycle adjacency graph for each primal-dual pair is drawn in blue over the corresponding bipartite graph representation in (a3), (b3), and (c3).}
  \label{fig:forbidden2}
\end{figure}

Here we prove \Cref{thm:minimal} from the main paper which we restate below.

\begin{theorem} \label{thm:minimal2}
  (Theorem 3 from the main paper) A cycle in $A_{(\C4)}$ defined by the sequence \mbox{$F=\langle C_1,C_2,\dots,C_k \rangle \subseteq\C_4$} for some tight cycle basis $\C$ of $T$ contains a common primal or dual node $x\in V(G_T)$ within each of the basis cycles $C_i\in F$ iff $F$ corresponds to a forbidden sub-hypergraph in $\HG$.
\end{theorem}

\begin{figure*}[tbp]
  \centering
  \subfloat[][Primal hypergraph with hyperedge simplifications applied]{\includegraphics[width=\columnwidth]{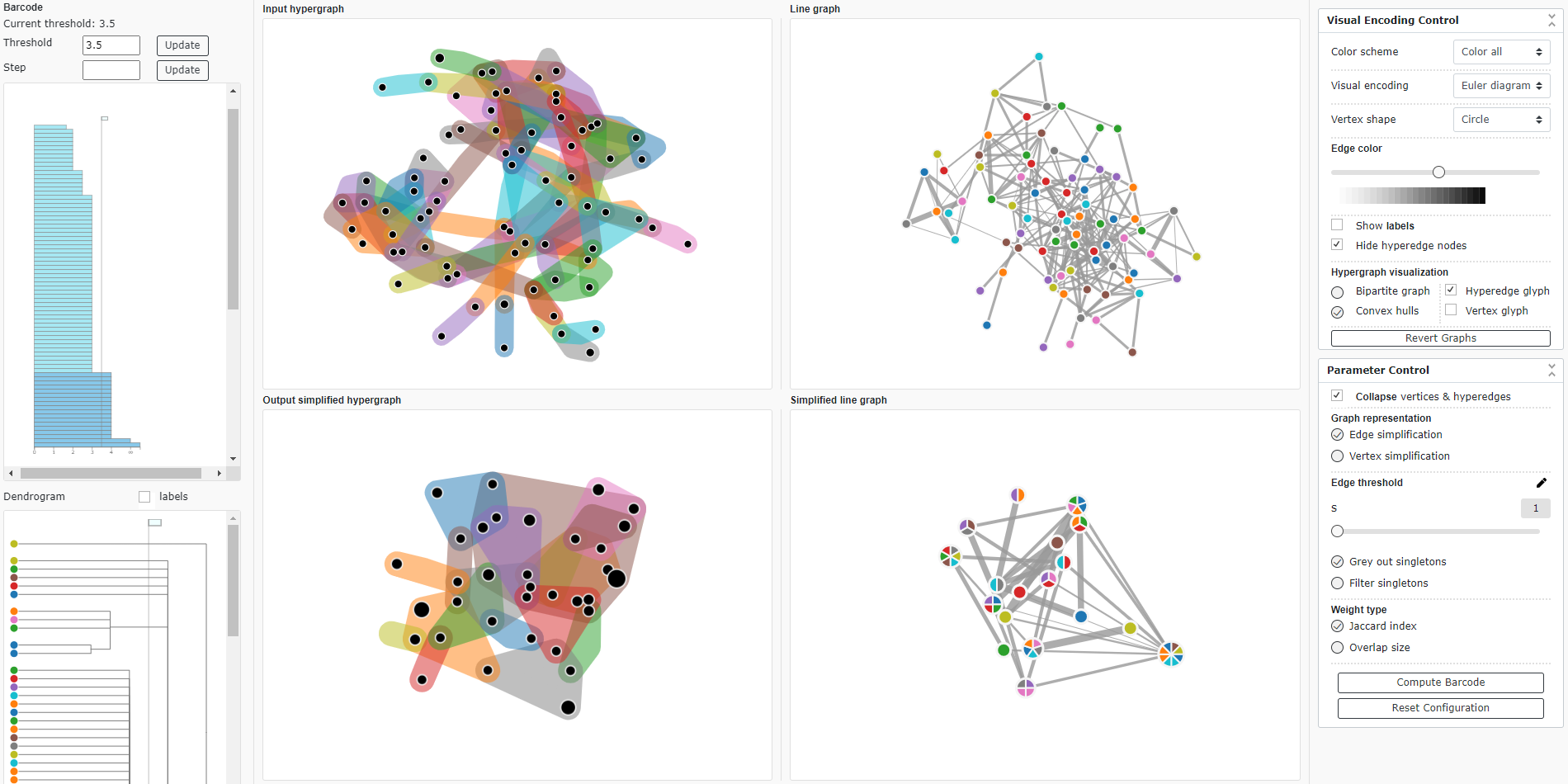}} \hspace{0.1in}
  \subfloat[][Dual hypergraph with vertex simplifications applied]{\includegraphics[width=\columnwidth]{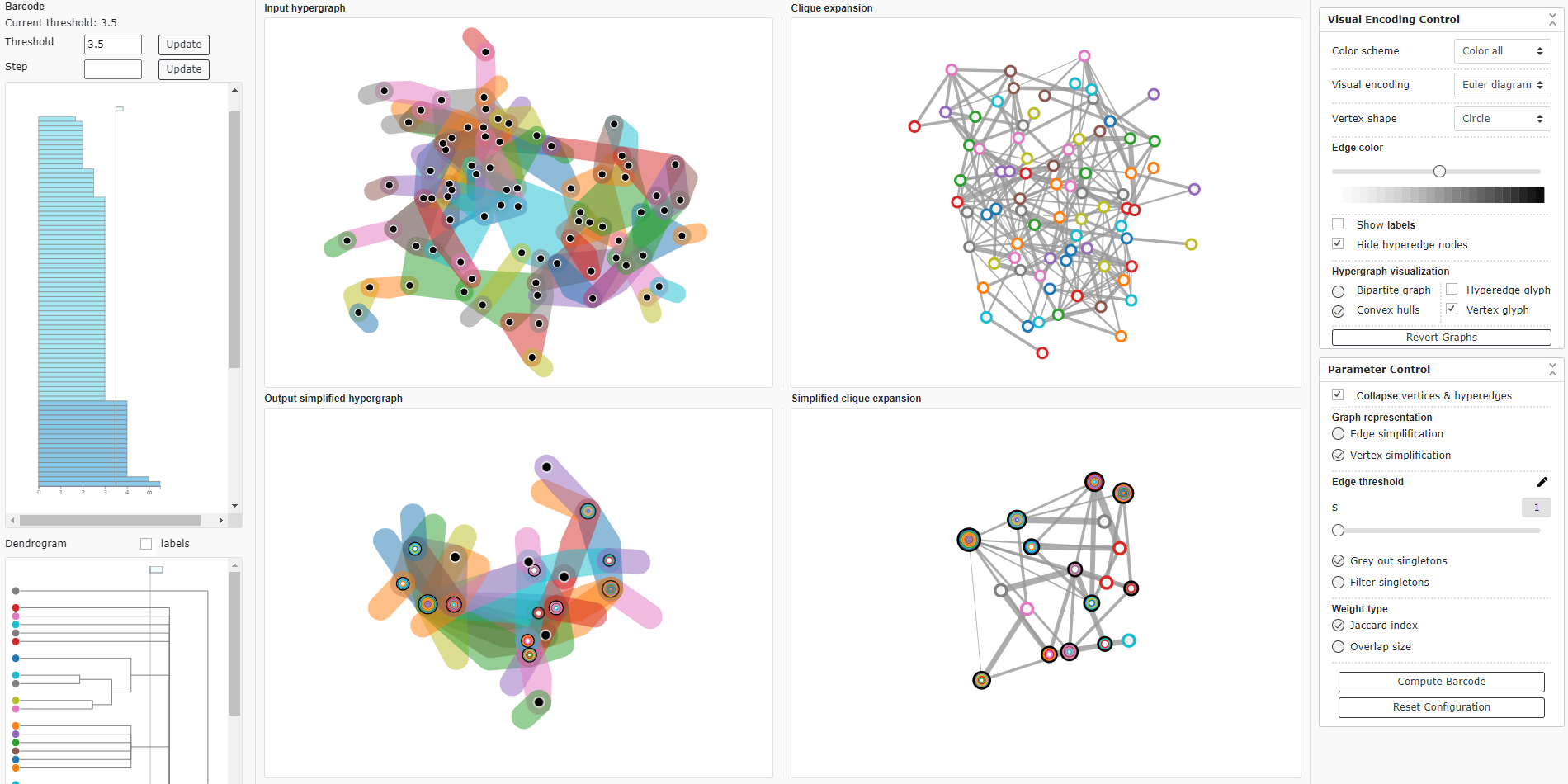}} \\
  \subfloat[][Primal hypergraph with vertex simplifications applied]{\includegraphics[width=\columnwidth]{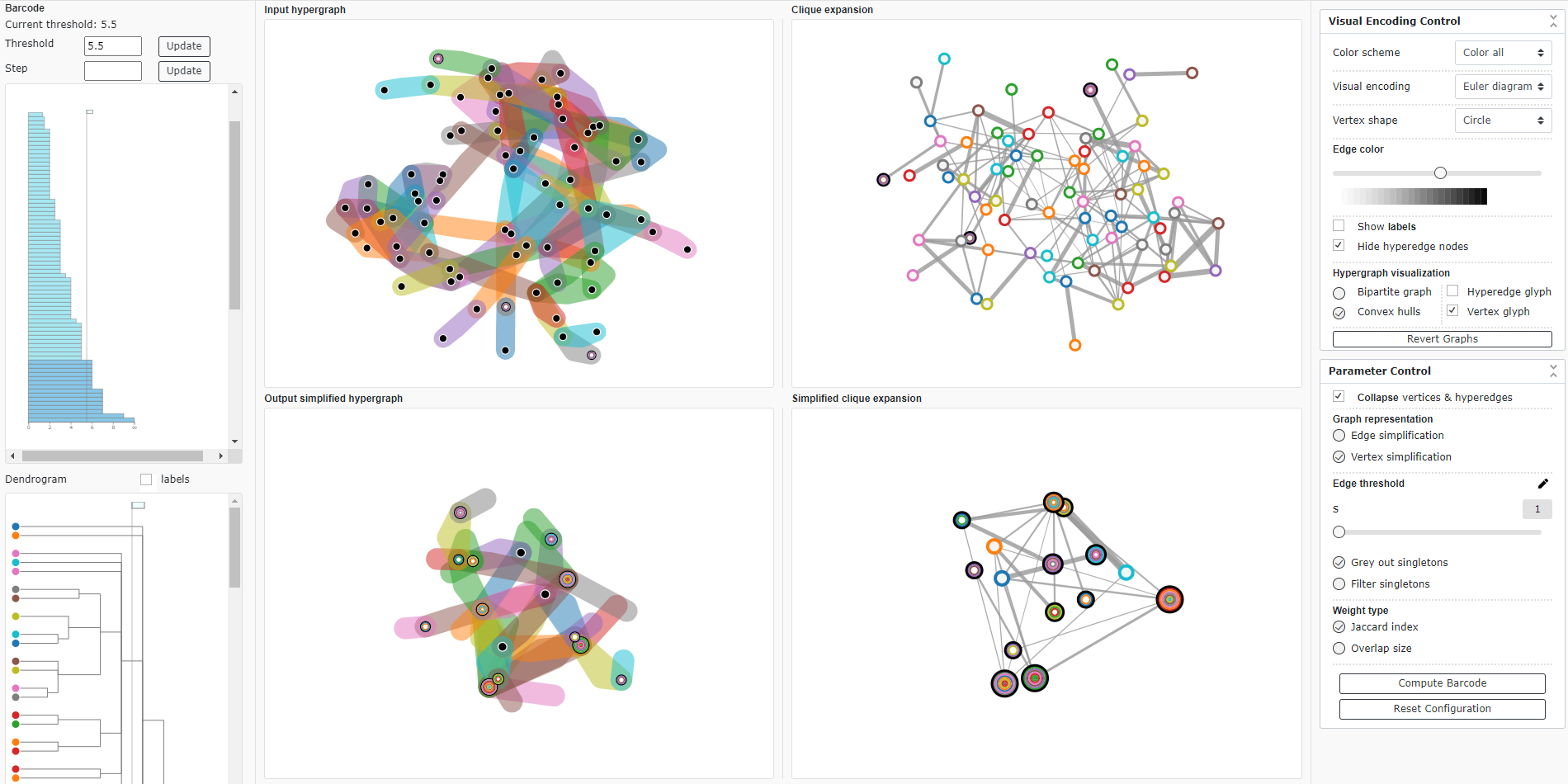}} \hspace{0.1in}
  \subfloat[][Dual hypergraph with hyperedge simplifications applied]{\includegraphics[width=\columnwidth]{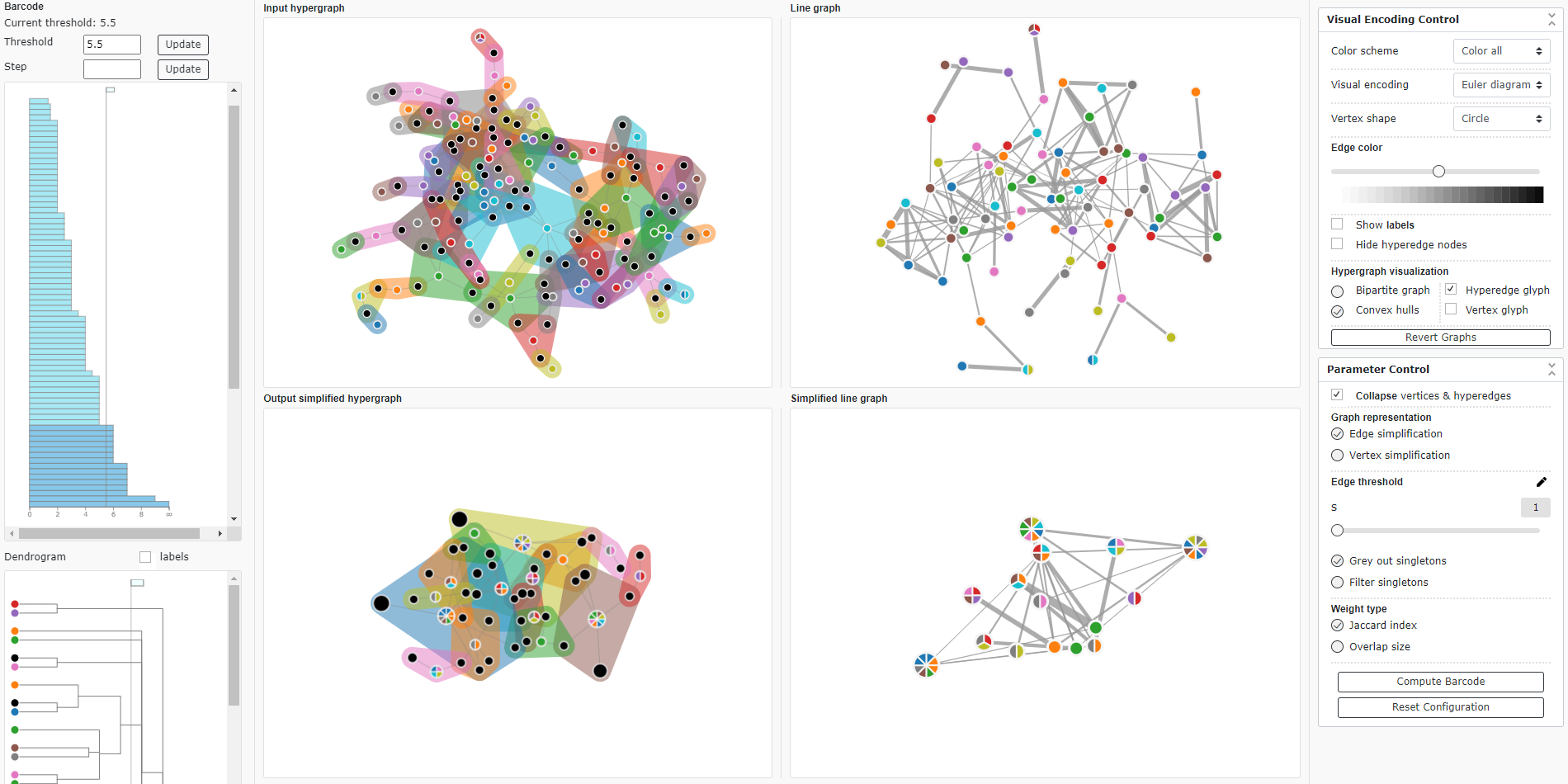}}
  \caption{Screen captures from the interactive tool presented in Zhou \etal~\cite{zhou2023simplification} with both the primal and dual hypergraphs of the museum interactions dataset from \cite{isella2011s}. Inside each screen capture, the input hypergraph and graph representation are shown on the top row of visualizations, the persistence barcode with simplification threshold applied on the left, and the simplified hypergraph and graph representation on the bottom row of visualizations.}
  \label{fig:zhou_interface}
\end{figure*}

\begin{proof}
  $\Leftarrow$ Since the basis cycles in $\C$ are linearly independent, a subset of minimal basis cycles, all having length 4, can have at most 3 common elements. Let $S$ denote the set of common elements for each $C_i \in F$ and assume that $|S|\geq 1$. We show that for any possible combination of elements in $S$, $F$ contains one of the forbidden sub-hypergraphs.

  In the case that $|S|=1$, the sole element in $S$ is either a primal or dual node of $G_T$. Without loss of generality, suppose that the sole element $v_0 \in S$ is a primal node corresponding to a vertex in the primal hypergraph. For $F$ to form a cycle in $A(\C_4)$, it must be that each consecutive pair of basis cycles $C_i,C_{i+1} \in F$ share an edge $(v_0,e_i)$ in $G_T$ where each $e_i\in V(G_T)$ is a dual node. Then each cycle $C_i \in F$ must contain elements $\langle e_{i-1},v_0,e_i,v_i \rangle$ where $v_i\in V(G_T)$ is a primal node not equal to $v_0$. It follows that the sequence $\langle v_1,e_1,v_2,e_2,\dots,v_k,e_k,v_1 \rangle$ defines a cycle in $T$. Since this sequence only contains hypergraph elements that are incident and adjacent to the primal vertex $v_0$, $v_0$ matches the definition of the cycle variant of a strangled vertex (\Cref{fig:forbidden2} (b1,b3)). Similarly, if the sole element in $e_0\in S$ is a dual node of $G_T$, corresponding to a hyperedge in the primal hypergraph, we can show that $e_0$ matches the definition of the cycle variant of a strangled hyperedge (\Cref{fig:forbidden2} (b2,b3)).

  Now consider the case where $S$ contains exactly one primal node $v_0\in V(G_T)$ and one dual node $e_0\in V(G_T)$. For $F$ to form a cycle in $A(\C_4)$, it must be that $|F| = k \geq 3$ since two cycles $C_i,C_j \in F$ sharing the edge $(v_0,e_0)$ in $G_T$ corresponds to a single edge in $A(\C_4)$ which is not a cycle. Then each basis cycle $C_i \in F$ must also contain primal and dual nodes $v_i,e_i\in E(G_T)$ where $v_i$ and $e_i$ are incident to each other, $v_i$ is incident to $e_0$, and $e_i$ is incident to $v_0$. It follows that the set $\{e_0,v_1,e_1,\dots,v_k,e_k\}$ induces a star sub-hypergraph in $T$ where $e_0$ is the central element and each sequence $\langle e_0,v_i,e_i\rangle$ for $i \geq 1 \leq k$, $k\geq3$, defines a point of the star. Similarly, the set $\{v_0,e_1,v_1,\dots,e_k,v_k\}$ induces a star sub-hypergraph centered on $v_0$ where each sequence $\langle v_0,e_i,v_i \rangle$ defines a point of the star. Since both of these sets only contain hypergraph elements incident and adjacent to the primal vertex $v_0$ and primal hyperedge $e_0$ respectively, $v_0$ matches the definition of the star variant of a strangled vertex, and $e_0$ matches the definition of the star variant of a strangled hyperedge (\Cref{fig:forbidden2}, (c1,c2,c3)).

  Finally, consider the case where $|S|\geq2$ and $S$ contains either a pair of primal nodes or a pair of dual nodes in $G_T$. Without loss of generality, suppose that $S$ contains a pair of primal nodes $v_0,v_1\in\V$. Then for $F$ to form a cycle in $A(\C_4)$, each consecutive pair of basis cycles $C_i,C_{i+1} \in F$ must share a dual node $e_i\in V(G_T)$ as well as edges $(v_0,e_i),(v_1,e_i)$. This means that each pair of consecutive basis cycles is connected by two edges in $A(\C_4)$, and each sub-sequence $\langle C_i,C_{i+1} \rangle$ is also a cycle in $A(\C_4)$. Considering only one of these sub-sequences, we have $C_i=\langle v_0,e_i,v_1,e_\alpha \rangle$ and $C_{i+1} = \langle v_0,e_i,v_1,e_\beta\rangle$ where $e_\alpha, e_\beta \in \E$ are distinct. Then $e_i,e_\alpha,e_\beta$ are all adjacent hyperedges with two common elements $v_0,v_1$, matching the definition for a 2-adjacent bundle of 3 hyperedges. We can further show that the original sequence $F$ contains elements matching the definition of a 2-adjacent bundle of $k+1$ hyperedges. If $S$ instead contains a pair of hyperedges $e_0,e_1\in\E$, we can similarly show that each sub-sequence $\langle C_i,C_{i+1} \rangle$ contains elements matching the definition of a 3-adjacent bundle of 2 hyperedges and $F$ contains elements matching the definition of a $(k+1)$-adjacent bundle of 2 hyperedges.

  $\Rightarrow$ From \Cref{fig:forbidden2} (a3,b3,c3), it is straightforward to verify that we can construct a minimum cycle basis for each of the forbidden sub-hypergraphs that satisfies the conditions of \Cref{thm:minimal2}.
\end{proof}

\section{Comparison to Zhou \etal~\cite{zhou2023simplification}}
\label{apx:zhou_comparison}

Here we provide a comparison of our structure-aware hypergraph simplification framework to the persistent homology guided simplification framework of Zhou \etal~\cite{zhou2023simplification}.

Zhou \etal use a line graph or clique expansion graph as the computational representations for the input hypergraph $\HG$, applying edge collapse operations on the chosen graph representation to induce hyperedge collapse or vertex collapse operations in $\HG$ respectively. The line graph representation has a node set corresponding to the hyperedges of $\HG$ with edges representing non-zero intersection between the vertex sets of the corresponding pair of hyperedges. These edges are weighted using the reciprocal of either the size of the intersection between the pair of hyperedges, or the Jaccard index between their vertex sets. The clique expansion graph representation has a node set corresponding to the vertices of $\HG$ with edges representing containment of the pair of corresponding vertices in at least one common hyperedge. Again, the edges are weighted using the reciprocal of either the number of common hyperedges or the Jaccard index of the containing hyperedge sets for the corresponding pair of vertices. They then apply persistent homology to a metric space of the chosen graph representation $G$ to generate a barcode where each bar represents an edge in a minimum spanning tree (MST) of $G$. The length of the bars corresponds to the edge weights. They collapse edges in the MST $G$ if the length of the corresponding bar in the barcode is less than a user controllable threshold, thereby inducing hyperedge or vertex simplifications in $\HG$.

By using separate graph representations, their approach is not able to generate simplified results that include both hyperedge and vertex simplifications simultaneously. Furthermore, the line graph and clique expansion graph representations are different depending on whether the primal hypergraph $\HG$ or dual hypergraph $\HG^*$ is used. In fact, the line graph of $\HG$ is equivalent to the clique expansion graph of $\HG^*$ and vice versa. By using the bipartite graph as our computational representation of $\HG$, which treats vertices and hyperedges in the same way, we naturally include hyperedge and vertex simplifications in a single unified framework. As discussed in our main paper, the bipartite graph representation $G(\HG)$ is equivalent to $G(\HG^*)$ and can be regarded as the same graph. Thus, our approach treats the primal and dual hypergraphs with equal importance while Zhou \etal~\cite{zhou2023simplification} may introduce inconsistency depending on the input and which graph representation is chosen.

\begin{figure*}
  \centering
  \subfloat[][Input hypergraph primal and dual \\ (157 total elements)]{\includegraphics[height=1.5in]{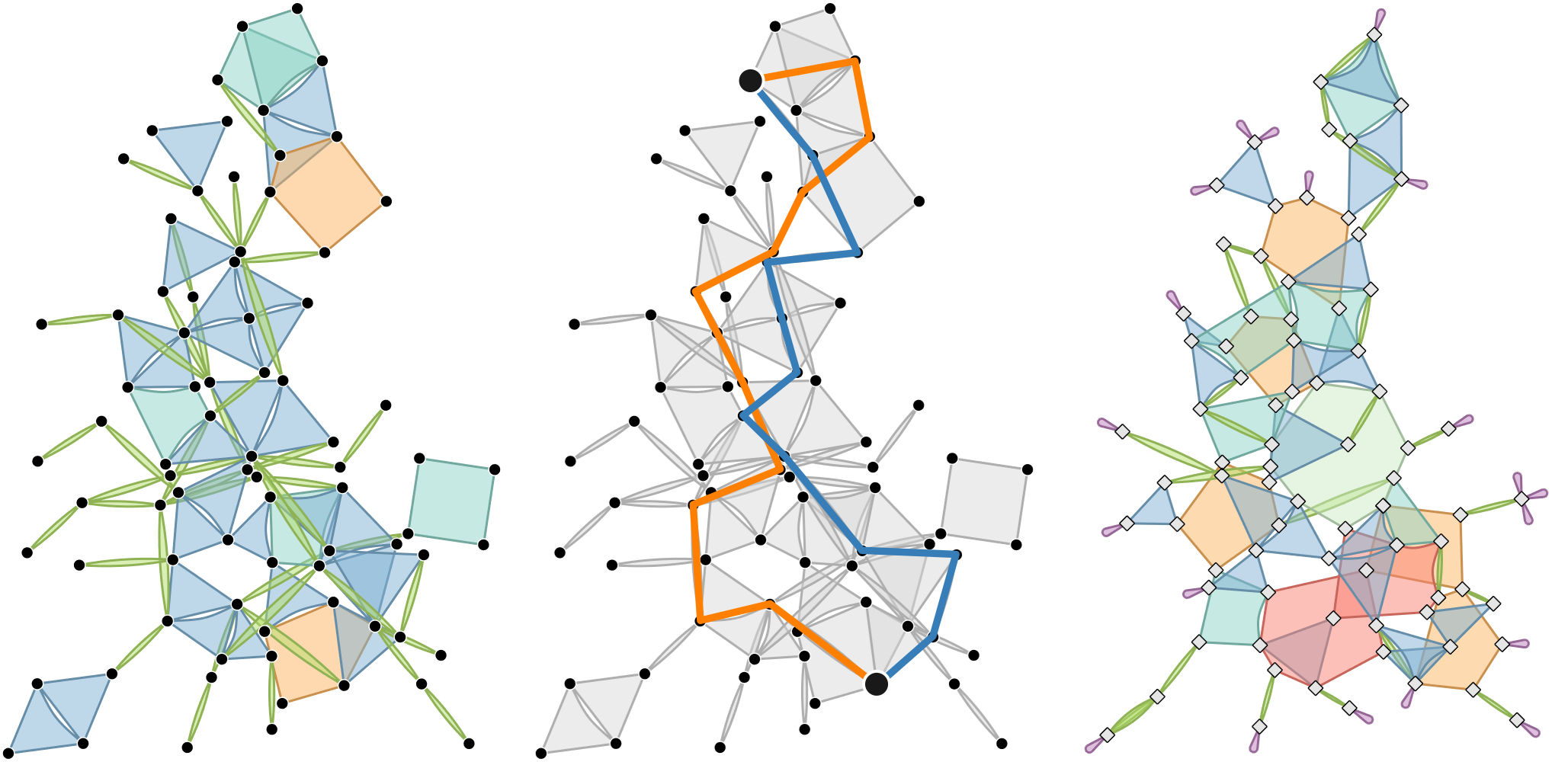}} \hspace{0.75in}
  \subfloat[][Our simplification primal and dual \\ (156 total elements)]{\includegraphics[height=1.5in]{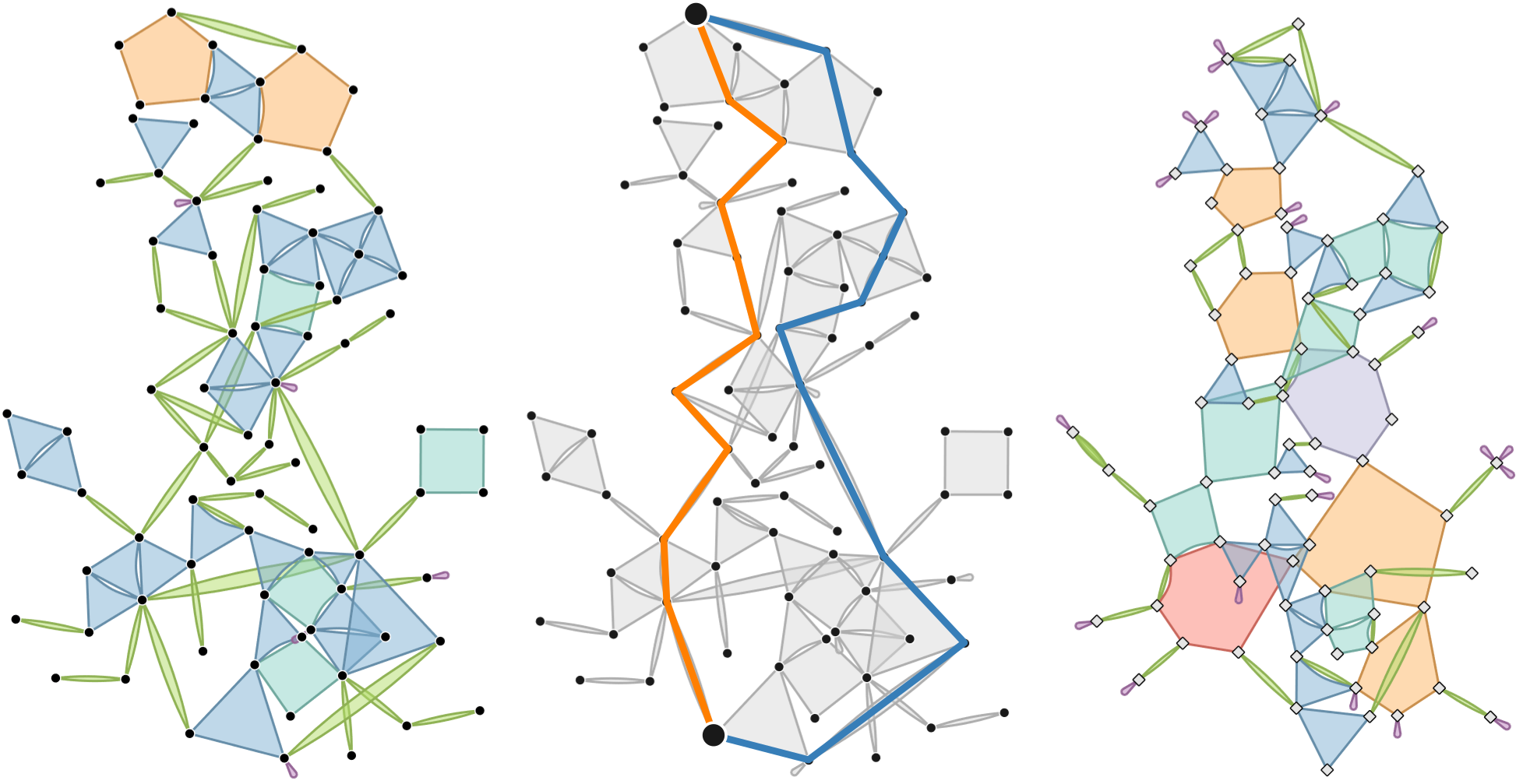}} \\ \vspace{0.0in} \hspace{-0.25in}
  \subfloat[][Zhou \etal hyperedge simplification in primal and vertex simplification in dual \\ (52 total elements)]{\includegraphics[height=1.25in]{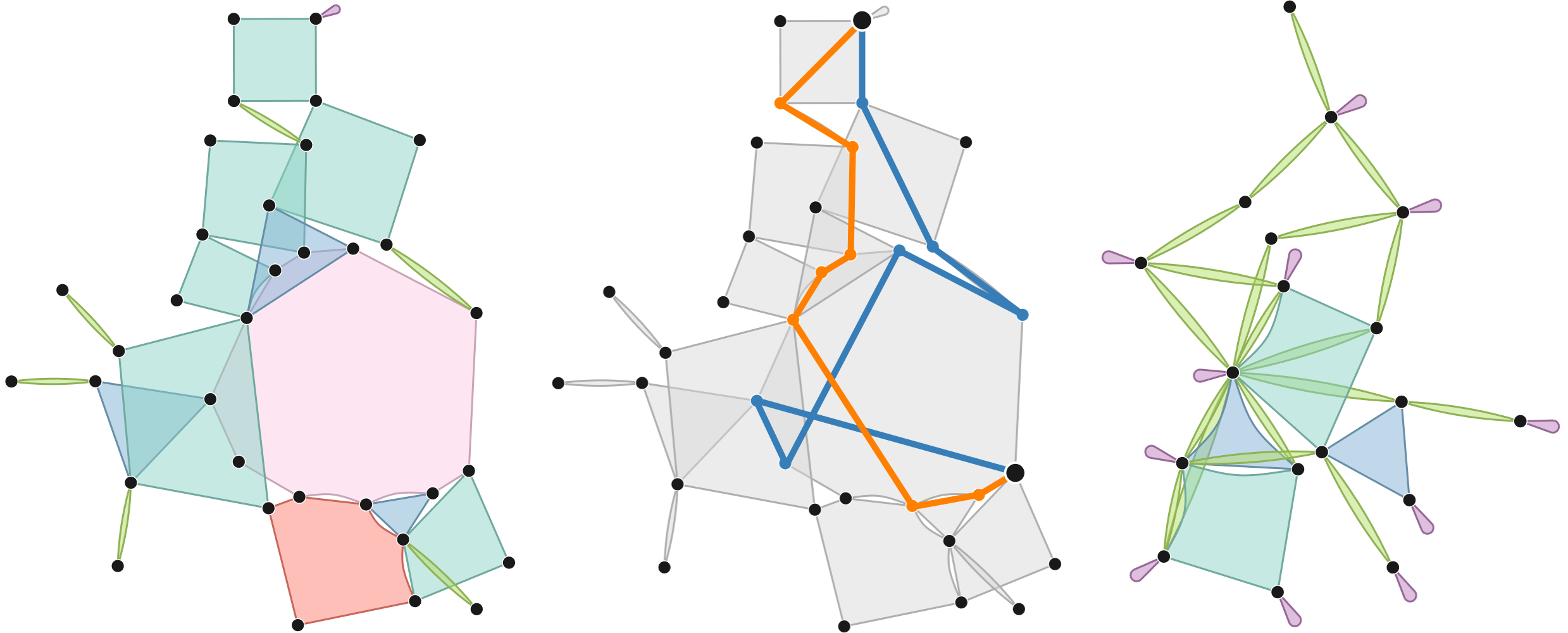}} \hspace{1.0in}
  \subfloat[][Zhou \etal vertex simplification in primal and hyperedge simplification in dual \\ (50 total elements)]{\includegraphics[height=1.25in]{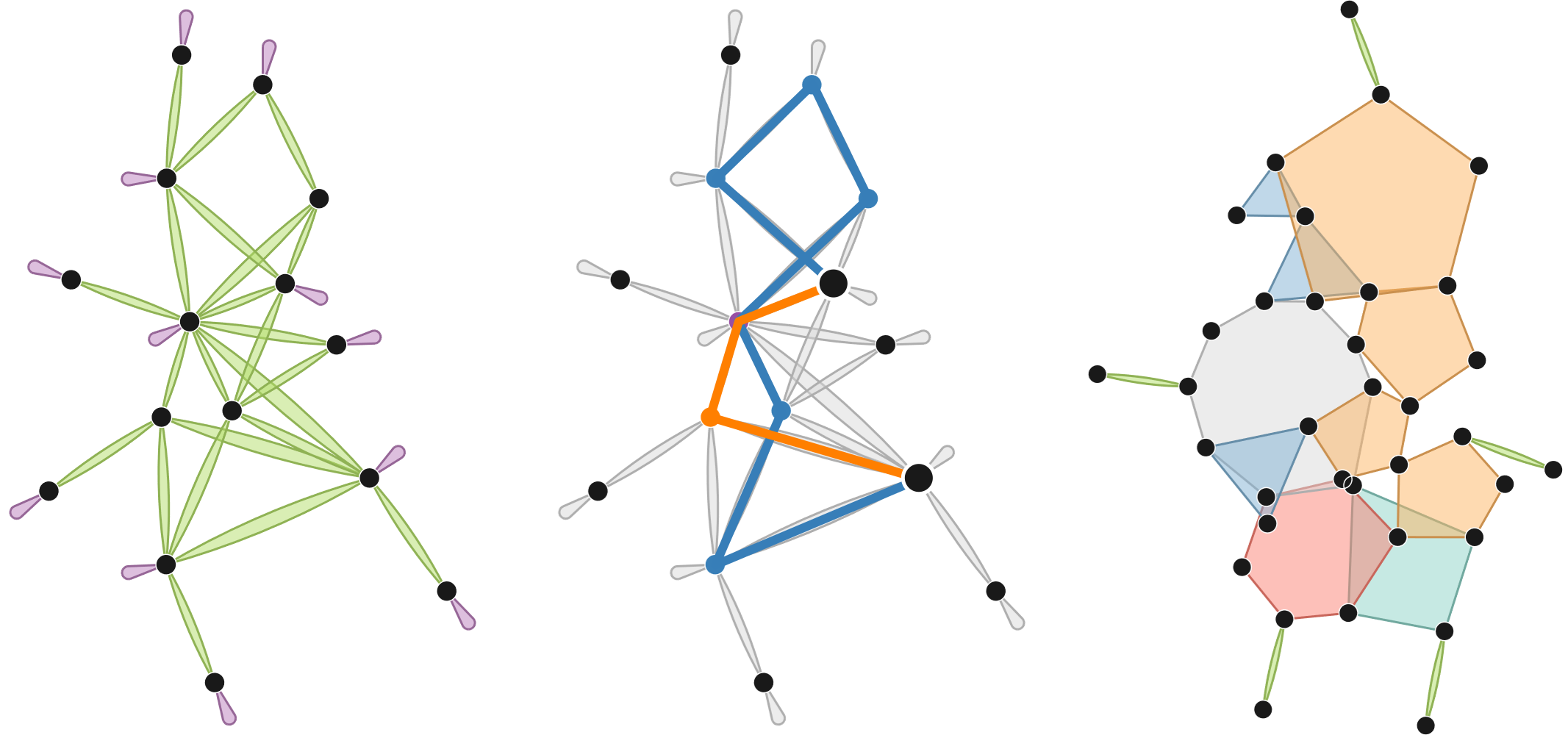}}
  \caption{A comparison of simplification results from Zhou \etal~\cite{zhou2023simplification} and ours for the museum interactions dataset from \cite{isella2011s}. Each sub-figure includes the primal and dual hypergraphs with polygons colored according to hyperedge cardinality on the left and right. In the middle, the same pair of hypergraph paths are highlighted in the primal hypergraph of each sub-figure indicated by orange and blue lines. These paths are disjoint in the original hypergraph (a) except for the start and end points.}
  \label{fig:zhou_compare}
\end{figure*}

\Cref{fig:zhou_interface} shows the museum interactions hypergraph, from \Cref{sec:results} of the main paper obtained from the dataset of Isella \etal~\cite{isella2011s}, loaded into the interactive tool presented by Zhou \etal~\cite{zhou2023simplification}. We accessed the tool from their GitHub repository at \url{https://github.com/tdavislab/Hypergraph-Vis/tree/master}. We tested their simplification method using both the hyperedge and vertex simplification options applied to both the primal hypergraph $\HG$ and dual hypergraph $\HG^*$ of the original dataset, visualizing the results in the default Euler diagram style. We adjusted the simplification threshold in all four cases to find the smallest value that would produce a Zykov planar hypergraph, \ie a hypergraph whose bipartite representation is a planar graph. We note that there are no additional layout options for the Euler diagram visualizations other than the default output of their system. This makes it difficult to observe any structures in the data, like cycles, bridges, and branches unless the layout is manually adjusted. However, it is convenient that the input hypergraph and simplified result are displayed next to each other.

Since the line graph representation of $\HG$ is equivalent to the clique expansion graph of $\HG^*$ and vice versa, the vertex simplified result in \Cref{fig:zhou_interface} (b) is the dual hypergraph of the hyperedge simplified result in (a), and the hyperedge simplified result in (d) is the dual hypergraph of the vertex simplified result in (c). This is also indicated by the fact that the barcode of (a) matches the barcode of (b) and the barcode of (c) matches the barcode of (d). However, since the primal and dual simplifications can only be performed separately, it is not possible to coordinate the layouts of the results, which could lead to different interpretations of the data and its structural features depending on which is used. More specifically, the upper left sub-windows in \Cref{fig:zhou_interface} (a) and (b) provide seemingly unrelated input visualizations, even though one is based on the primal hypergraph and the other on the dual hypergraph of the same data. Furthermore, between \Cref{fig:zhou_interface} (a) and (c), even though the visualization of the input hypergraph (upper left sub-windows) is the same, the simplified results (the sub-windows below the input visualizations) are different and seemingly unrelated. This is due to the different types of simplification operations that are used: (a) hyperedge simplification based on the clique expansion graph representation, and (c) vertex simplification based on the line graph representation. In contrast, we use the bipartite graph representation to handle the primal hypergraph and its dual in a consistent fashion.

We also compare their simplifications to ours using the polygon visualization metaphor. We exported the simplified hypergraphs from Zhou et al.'s tool and imported them into our polygon visualization tool which is available at \url{https://github.com/tdavislab/Hypergraph-Vis/tree/master}. \Cref{fig:zhou_compare} compares these visualizations of the original museum interactions hypergraph (a), our simplified result from the main paper \Cref{fig:museum} (b), the hyperedge simplification of the primal hypergraph from Zhou \etals tool~\cite{zhou2023simplification} paired with their vertex simplification of the dual hypergraph (c), and the vertex simplification of the primal hypergraph from Zhou \etals tool (d) paired with their hyperedge simplification of the dual hypergraph, all using the polygon visualization metaphor. The dual hypergraph visualizations in (a) and (b) are coordinated with the primal visualizations such that corresponding primal and dual elements appear in approximately the same locations. Since they are generated separately, the dual visualizations in (c) and (d) are not coordinated with the primal visualizations, resulting in visualizations that may look unrelated. We observe that both the hyperedge and vertex simplification methods of Zhou \etal are effective in reducing the size of the data while keeping some of the original features, but much of the data is simplified away to reach a result that is Zykov planar, keeping only 50--52 elements compared to the original 157. Our result, on the other hand, keeps 156 elements, retaining most of the original information and also preserving the path structures indicated by the orange and blue highlights in the middle visualization of each sub-figure. These highlights represent a pair of paths that are disjoint in the original hypergraph \Cref{fig:zhou_compare} (a) except at the starting and ending vertices. In (b), we can clearly see that these paths are still distinct. In (c) however, the paths intersect through a common super-hyperedge, and in (d), they intersect through a common super-vertex. Additionally, the hyperedge and vertex simplification methods alter the paths in different ways as we can see with the orange path in (d) which is significantly shorter than the orange path in (c). For a visualization application that requires identifying disjoint paths, such as identifying transmission vectors of an infectious disease, we argue that our simplification provides a benefit over Zhou \etals.

A more thorough comparison of Zhou \etals~\cite{zhou2023simplification} with ours is challenging because of the difference in simplification goals. We aim to reduce clutter by reducing unavoidable overlaps while preserving structural information. Zhou \etal aim to reduce visual clutter by reaching a compact representation that can be used for analysis. Both approaches have applications for which they may be more appropriate. Additionally, we cannot directly compare results based on the number of simplifications applied or the number of elements in the simplified results because the frameworks have different termination criteria. Ours is based on the removal of unavoidable overlaps and theirs is based on a persistence threshold value, \ie if a set of features have the same barcode length, they are either all simplified or all not simplified.

\section{Enlarged Figures}
\label{apx:figures}

On the following pages, we provide enlarged versions of \Cref{fig:paperauthor} and \Cref{fig:museum} from the main paper.

\begin{figure*}[h]
  \centering
  \subfloat[][Input hypergraph]{\includegraphics[height=3.75in]{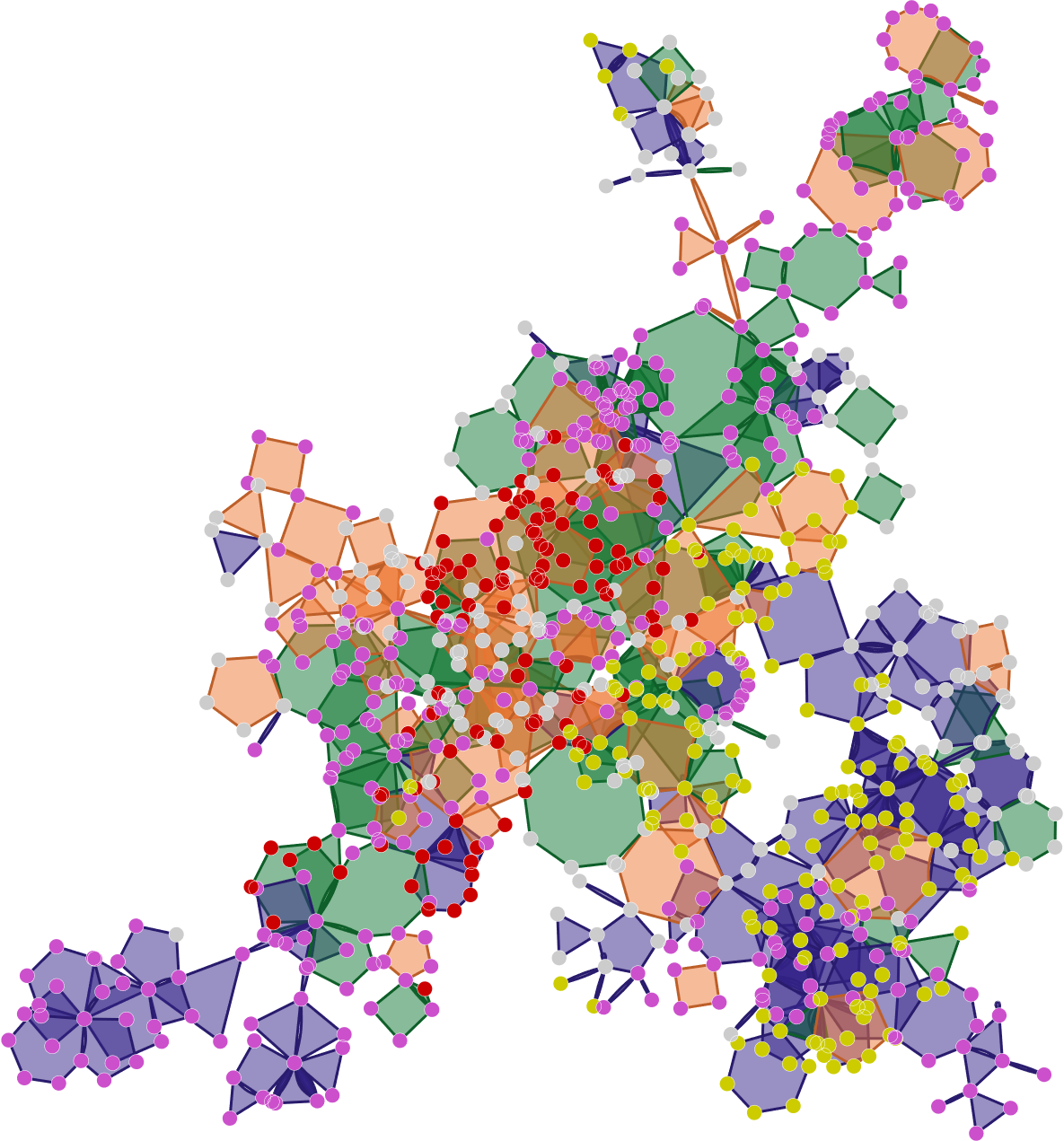}} \hspace{0.25in}
  \subfloat[][Oliver \etal simplified scale]{\includegraphics[height=3.75in]{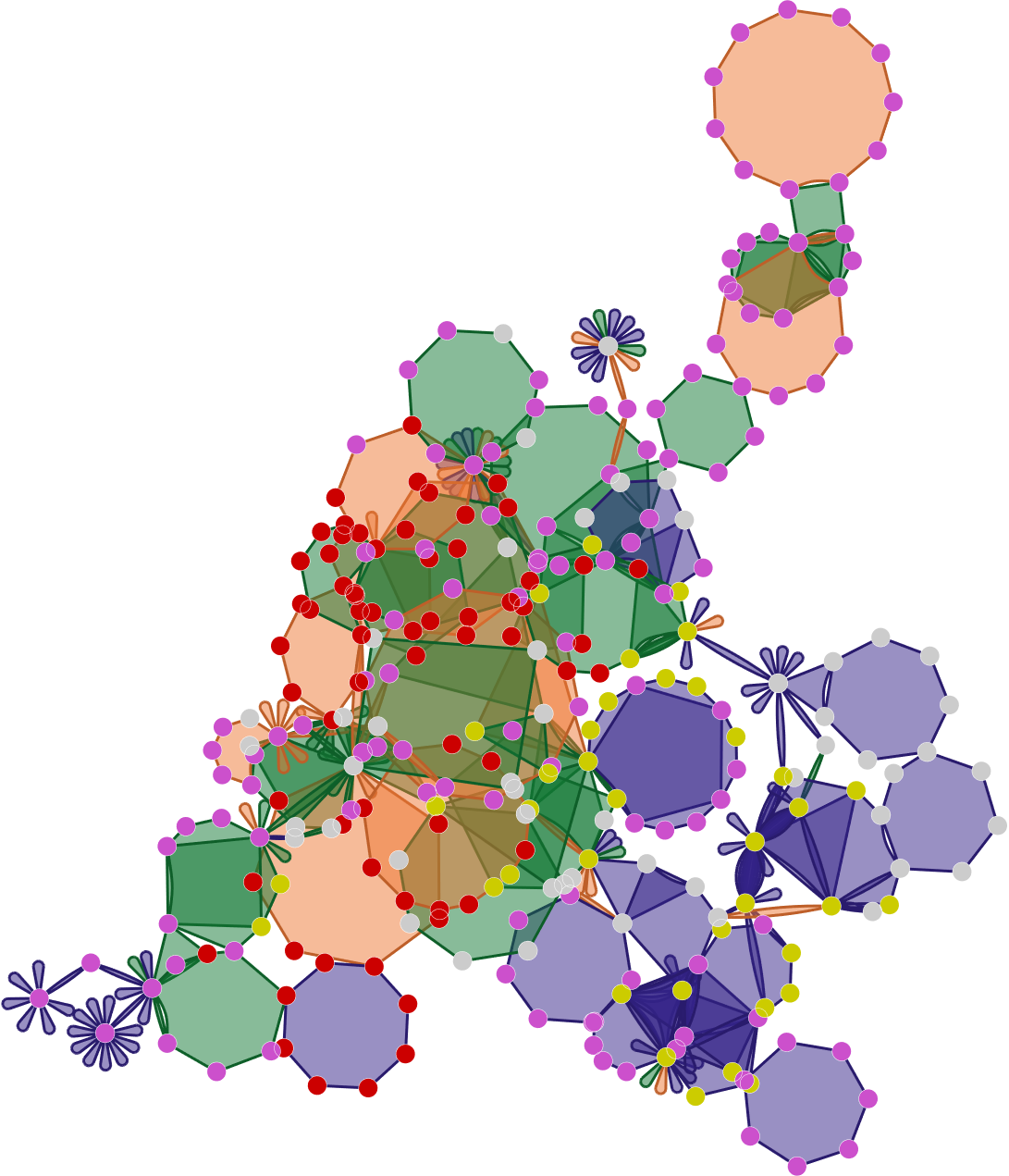}} \\ \vspace{0.5in}
  \raisebox{1.75in}{\includegraphics[height=2in]{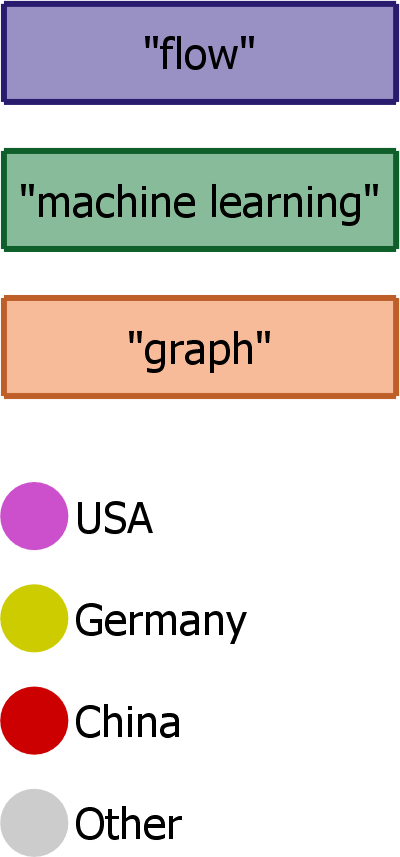}}
  \subfloat[][Our simplified scale]{\includegraphics[height=3.75in]{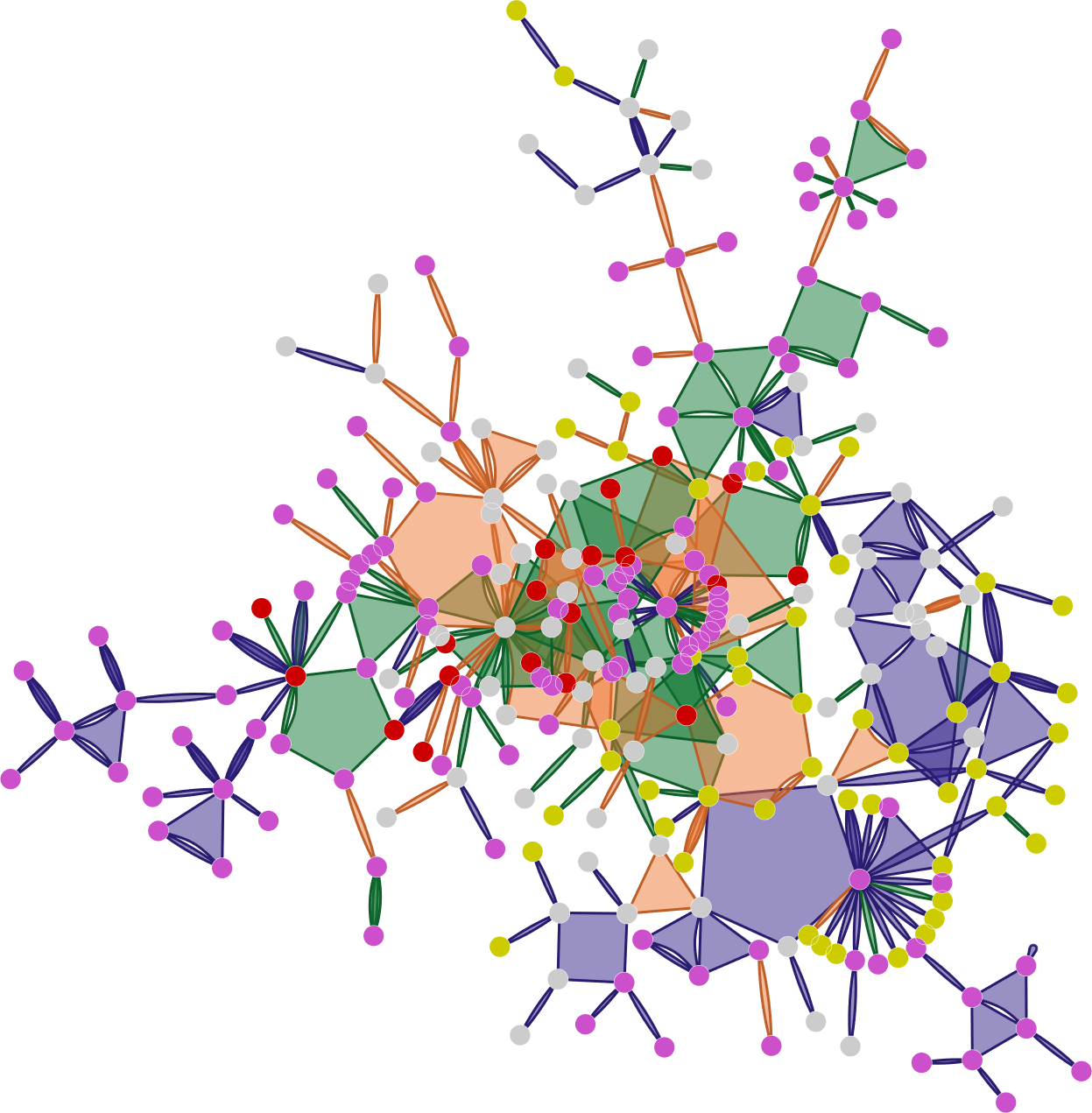}}
  \caption{Enlarged versions of the visualizations in \Cref{fig:paperauthor} from the main paper. A paper-author hypergraph network with 786 vertices and 318 hyperedges, (a) is simplified using the priority guided approach of Oliver \etal~\cite{oliver2024scalable} (b), and our topological decomposition guided approach (c). Notice that some of the branches in (b) have been reduced and others have been eliminated entirely. Our hypergraph decomposition extracts structures ahead of time, allowing us to preserve the skeleton of each branch. In each visualization, the hyperedges are colored according to the keywords of their papers: blue for the keyword ``flow'', green for the keywords ``machine learning'', and orange for the keyword ``graph''. Additionally, the vertices are colored according to the geographic location of the affiliated research institution for each author: magenta for institutions based in the United States, yellow for institutions based in Germany, red for institutions based in China, and gray for institutions based elsewhere.}
  \label{fig:paperauthor2}
\end{figure*}

\begin{figure*}[h]
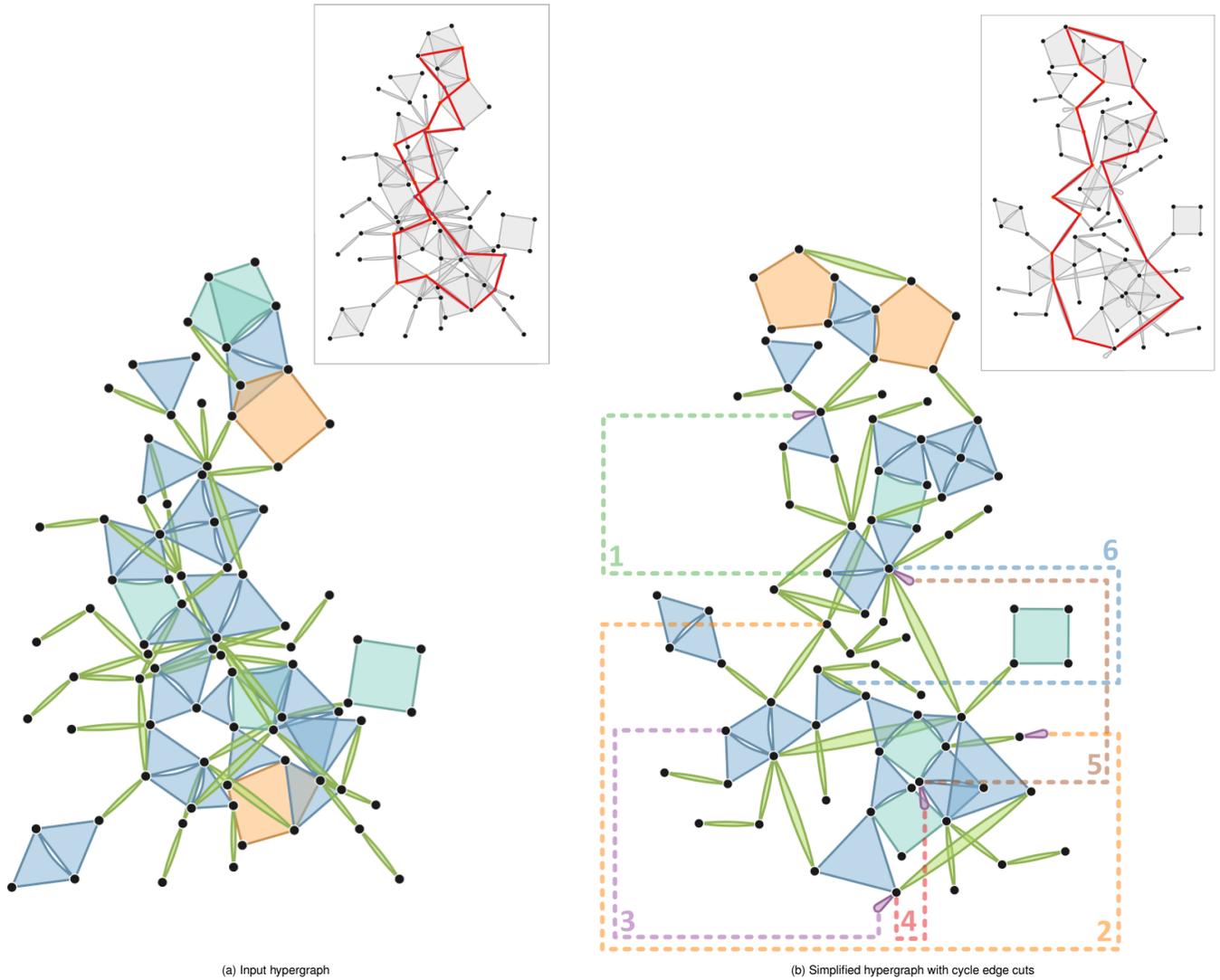

  \centering
  \subfloat[][Input hypergraph]{\includegraphics[height=5.5in]{museum_nosimp_paths.png}} \hspace{0.25in}
  \subfloat[][Simplified hypergraph with cycle edge cuts]{\includegraphics[height=5.5in]{museum_cutsimp_paths.png}}
  \caption{Enlarged versions of the visualizations in \Cref{fig:museum} from the main paper. A hypergraph representing face-to-face interactions between museum visitors \cite{isella2011s} is visualized before, (a), and after simplifying the topological blocks, (b). Notice the numerous triangles and digons in (a) that overlap despite being non-adjacent. On the right, six cycles have been cut to make the hypergraph planar. The deleted incidence relationships between vertices and hyperedges are represented by dashed annotation lines along the boundary of the visualization. In the upper right corners, we highlight the same two paths for each layout from a vertex near the top to a vertex near the bottom.}
  \label{fig:museum2}
\end{figure*}

\end{document}